\documentclass[lettersize,journal]{IEEEtran}
\usepackage{amsmath,amsfonts}
\usepackage{algorithmic}
\usepackage{algorithm}
\usepackage{array}
\usepackage[caption=false,font=normalsize,labelfont=sf,textfont=sf]{subfig}
\usepackage{textcomp}
\usepackage{stfloats}
\usepackage{booktabs}
\usepackage{multirow}
\usepackage{url}
\usepackage{verbatim}
\usepackage{graphicx}
\usepackage{cite}
\usepackage{multicol}
\usepackage{makecell}
\usepackage{svg}
\begin{document}

\title{Synesthesia of Machines (SoM)-Aided LiDAR Point Cloud Transmission for Collaborative Perception}

\author{Ensong Liu,~\IEEEmembership{Graduate Student Member,~IEEE}, Rongqing Zhang,~\IEEEmembership{Member,~IEEE}, \\ Xiang Cheng,~\IEEEmembership{Fellow,~IEEE} and Jian Tang,~\IEEEmembership{Fellow,~IEEE}
\thanks{Ensong Liu and Xiang Cheng are with the State Key Laboratory of Advanced Optical Communication Systems and Networks, School of
 Electronics, Peking University, Beijing 100871, P. R. China (e-mail: ensongliu@pku.edu.cn; xiangcheng@pku.edu.cn).}
\thanks{Rongqing Zhang is with the School of Software Engineering, Tongji University, Shanghai 200092, P. R. China (e-mail: rongqingz@tongji.edu.cn).}
\thanks{Jian Tang is with Beijing Innovation
Center of Humanoid Robotics, Beijing 101111, P. R. China (e-mail: jian.tang@x-humanoid.com).}}

\markboth{Journal of \LaTeX\ Class Files,~Vol.~14, No.~8, August~2021}%
{Shell \MakeLowercase{\textit{et al.}}: A Sample Article Using IEEEtran.cls for IEEE Journals}


\maketitle

\begin{abstract}
Collaborative perception enables more accurate and comprehensive scene understanding by learning how to share information between agents, with LiDAR point clouds providing essential precise spatial data. Due to the substantial data volume generated by LiDAR sensors, efficient point cloud transmission is essential for low-latency multi-agent collaboration. In this work, we propose an efficient, robust and applicable LiDAR point cloud transmission system via the Synesthesia of Machines (SoM), termed LiDAR Point Cloud Feature Transmission (LPC-FT), to support collaborative perception among multiple agents. Specifically, we employ a density-preserving deep point cloud compression method that encodes the complete point cloud into a downsampled efficient representation. To mitigate the effects of the wireless channel, we design a channel encoder module based on self-attention to enhance LiDAR point cloud features and a feature fusion module based on cross-attention to integrate features from transceivers. Furthermore, we utilize the nonlinear activation layer and transfer learning to improve the training of deep neural networks in the presence the digital channel noise. Experimental results demonstrate that the proposed LPC-FT is more robust and effective than traditional octree-based compression followed by channel coding, and outperforms state-of-the-art deep learning-based compression techniques and existing semantic communication methods, reducing the Chamfer Distance by 30\% and improving the PSNR by 1.9 dB on average. Owing to its superior reconstruction performance and robustness against channel variations, LPC-FT is expected to support collaborative perception tasks.
\end{abstract}


\section{Introduction}
\IEEEPARstart{P}{erception} is a significant module of an autonomous driving system, which utilizes sensors to monitor and understand the surrounding environments. Although individual perception has been extensively studied with the development of deep learning in recent years, it suffers from the occlusion issues stemmed from individual limited line-of-sight visibility. To address these issues, collaborative perception exploits the interaction among multiple agents, wherein perceptual data from several nearby agents are shared to achieve a more comprehensive and accurate understanding of the environment \cite{han2023collaborative}. Light detection and ranging (LiDAR) is a pivotal sensor technology for autonomous driving due to their high-resolution and long-range perception capabilities \cite{li2020lidar}. Leveraging LiDAR’s omnidirectional sensing capabilities and 3D environmental representation, collaborative perception can effectively overcome visual limitations of individual perception. However, perception tasks are latency-sensitive due to the high mobility of the traffic environment. The large data volume of raw LiDAR point clouds hinders the effective multi-agent collaboration perception. Therefore, it is imperative to develop innovative LiDAR-based collaborative perception systems, which can achieve both communication bandwidth efficiency and reliable perception capability. 

Collaborative perception can be categorized into early fusion\cite{chen2019cooper}, intermediate fusion\cite{li2021learning, xu2022v2x,wang2023core} and late fusion based on the level of information shared between agents. Among these, intermediate fusion has garnered the most attention in existing studies due to its favorable trade-off between performance and bandwidth efficiency. This approach combines compressed intermediate feature representations from multiple agents. However, current intermediate fusion works primarily focus on the object-level tasks, such as object detection \cite{li2021learning,xu2022v2x}. For other autonomous driving tasks, e.g, planning and prediction, agents need to understand the complete surrounding environment, including both foreground objects and background information. Consequently, it is crucial to develop novel methods for transmitting features that enable the reconstruction of complete LiDAR point clouds. These methods should efficiently compress point clouds while preserving environmental information as much as possible. Moreover, most studies on collaborative perception assume ideal transmission conditions for inter-agent communications. Although some studies have considered communication issues like latency \cite{xu2022v2x} and packet loss \cite{li2023learning}, the direct impact of wireless channels on collaborative perception remains an underexplored research area. In summary, the transmission of LiDAR point clouds imposes significant demands on communication bandwidth and is vulnerable to interference of wireless channels, thereby constraining the effectiveness of collaborative perception systems.

Fortunately, semantic communications is promising to overcome the above-mentioned issues, which can maintain high task performance subject to the limited communication condition. In contrast to conventional communications, semantic communications only transmit essential information relevant to the specific task at the receiver, resulting in a substantial reduction in data traffic \cite{qin2021semantic, yang2022semantic}. This object can be achieved via the Synesthesia of Machines (SoM), which aims to exploit compact, task-aware and robust features contained in original information, called Synesthesia of Machines-Feature (SoM-Feature), through designing communication and multi-modal sensing fusion neural networks \cite{cheng2023intelligent}. Therefore, semantic communications have the potential to facilitate the transmission of massive data in collaborative perception, ensuring real-time, high-performance system operation. Many existing works on semantic communications concentrate on tasks in natural language processing (NLP) and computer vision (CV), including text transmission \cite{xie2020lite, xie2021deep,getu2024performance} image reconstruction \cite{bourtsoulatze2019deep,xu2021wireless,yang2023witt,ma2023task} and image classification \cite{kutay2024classification,hua2024classification}. However, there is a lack of in-depth research on semantic communications for LiDAR point clouds. This gap limits the application potential of semantic communications in multi-agent collaborative perception. 

For multi-agent collaborative perception, different agents need to communicate self-perceived LiDAR point clouds with each other. The agents further construct the environment map utilizing the shared LiDAR point clouds, and conduct high-level perception tasks based on the constructed map. Therefore, the LiDAR point cloud transmission approach must meet the requirements for both real-time transmission and high-accuracy reconstruction. Our object is to design an efficient, robust, and practical LiDAR point cloud transmission system for multi-agent collaborative perception. Specifically, the following difficulties for achieving efficient LiDAR point cloud transmission should be addressed:
\begin{enumerate}
        \item Representing LiDAR point clouds in a memory-efficient manner poses a significant challenge due to their non-uniform nature and rich informational content, which is more complex compared to regular point clouds that describe the shape of a single object.
        \item Wireless channel fading causes severe distortion for transmitted LiDAR point cloud features. It is crucial to design a point cloud feature enhancement network that ensures accurate point cloud reconstruction despite the presence of wireless channel noise. 
        \item Most current semantic communication works assume the discrete-time analog transmission (DTAT) \cite{shao2022semantic}, treating continuous features as transmission symbols and neglecting the digital components of the practical wireless communication system.
\end{enumerate}
To address these challenges, we design a novel LiDAR point cloud feature transmission system (LPC-FT) for collaborative perception under the guidance of SoM \cite{cheng2023intelligent}. The transmitter first encodes the LiDAR point cloud into compact features using a density-preserving feature encoder and then enhances these features through a self-attention-based channel encoder network. These encoded features are then transmitted to over a wireless channel. At the receiver, a symmetric channel decoder network decodes the noisy features, which are subsequently fused with the receiver-perceived LiDAR point cloud using a cross-attention-based feature fusion network. Finally, an adaptive feature decoder network reconstructs the transmitted LiDAR point cloud. Our results demonstrate that the proposed LPC-FT system outperforms existing point cloud semantic communication methods, achieving a 30\% reduction in point-to-point distance on average and an improvement of 1.9 dB in point-to-plane PSNR. The main contributions of this paper are summarized as follows:
\begin{itemize}
        \item To support accurate and efficient collaborative perception, we propose a LiDAR point cloud feature transmission system (LPC-FT). The proposed system consists of four key components: feature encoder and decoder, channel encoder and decoder, feature fusion and a nonlinear actiavtion layer. We utilize a two-stage training strategy to optimize the whole network, which can leverage the pretrained knowledge of point cloud compression to boost reconstuction performance under various wireless channel conditions.
        \item Due to the unique challenges of LiDAR point clouds, we employ a density-preserving point cloud compresssion network as the feature encoder and decoder. This network effectively captures the local density of LiDAR point clouds and adaptively reconstructing point clouds. Additionally, to improve the reconstruction performance further, we incorprate absolute coordinate position encoding into the point transformer block accounting for the large-scale distribution of LiDAR point clouds.
        \item To ensure robust performance under the effects of wireless channel, we propose a point cloud channel encoder network based on self-attention to cope with wireless channel noise. Furthermore, we design a point cloud feature fusion network at the receiver, which exploits the correlations of the transmitted features and the receiver-perceived features based on cross-attention.
        \item Unlike existing analog semantic communication systems, our proposed LPC-FT system adopts a digital communication framework with quantization and modulation, ensuring compatibility with existing communication systems. To address the challenge of non-differentiable digital operations, we employ a nonlinear straight-through estimator (STE), enabling gradient propagation and mitigating the effects of digital quantization. This approach stabilizes network optimization and enhances system performance.
\end{itemize}

\begin{figure*}[t]
    \centering                                   
    \includegraphics[width=0.95\linewidth]{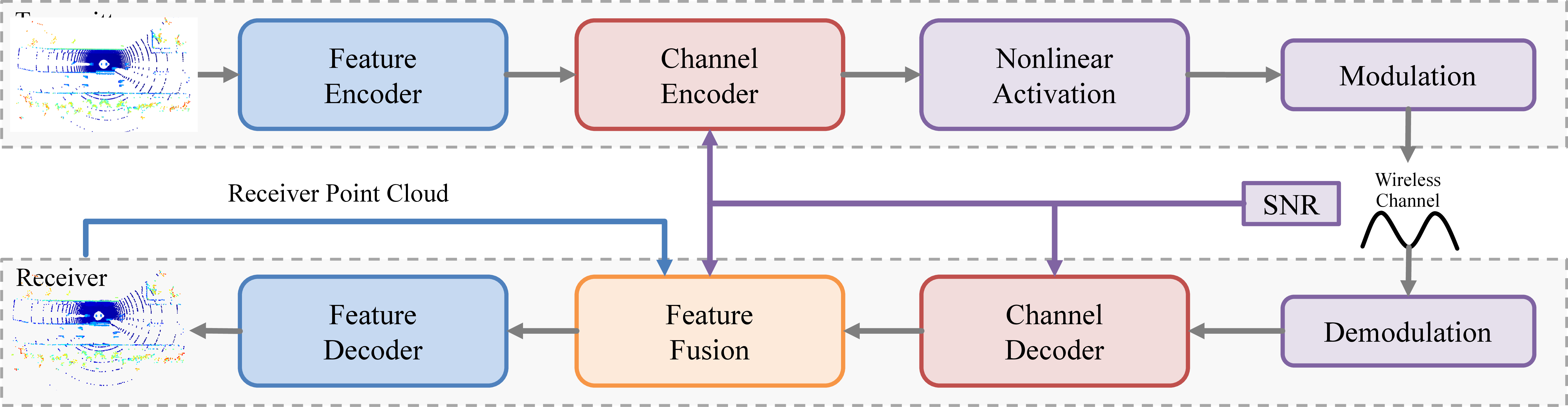}    
    \caption{Overview of the LPC-FT, our proposed LiDAR point cloud feature transmission system. The system have four key components: feature encoder and decoder, channel encoder and decoder, feature fusion, and nonlinear activation layer. }
    \label{img:overview} 
    \vspace{-0.5cm}
\end{figure*}

\section{Related Works}

\subsection{LiDAR Point Cloud Compression}
Many works in the field of robotics and perception focus on compressing point cloud data to support real-time task execution. Traditional point cloud compression algorithms \cite{galligan2018google,graziosi2020overview} usually rely on octree \cite{meagher1982geometric} or KD-tree \cite{bentley1975multidimensional} structure for storage efficiency. While tree strctures are very memory-efficient in most real-world scenarios, they often fail to capture structural details effectively. With the development of deep learning, learning-based point cloud compression has gained widespread concern with the benefit of extracting features efficiently. Learning-based point cloud compression usually applies the autoencoder network architecture, which utilizes the foundational layers for point cloud processing \cite{qi2017pointnet,qi2017pointnet++,zhao2021point} to encode point clouds into a compact representation\cite{huang20193d,zhang2022transformer}. With the advancement of LiDAR technology in autonomous driving, numerous studies have concentrated on the efficient compression of LiDAR point clouds. A novel deconvolution operator was proposed to decompress dense LiDAR point cloud maps from the compact point features \cite{wiesmann2021deep}. A deep learning framework employing the octree structure was proposed to solve the performance degradation of voxel-based methods\cite{fu2022octattention}. In \cite{he2022density}, the authors considered a novel point cloud compression network that preserves local density information, which can achieve the state-of-the-art compression performance for both object and LiDAR point clouds.

Given that local density is a crucial characteristic of LiDAR point clouds, we adopt the density-preserving point cloud compression method as described in \cite{he2022density} and add the extra absolute position encoding to realize better compression. Inspired by the attention mechanism for point clouds \cite{zhao2021point}, we further propose an attention-based channel encoder network and feature fusion network to combat against wireless channel effects and improve reconstruction quality.

\subsection{LiDAR Point Cloud Semantic Communications}
Weaver \cite{shannon1949mathematical} categorized communications into three levels: the technical level, the semantic level, and the effectiveness level. Semantic communications focus on the semantic level and the effectiveness level \cite{qin2021semantic}, transmiting only the necessary information relevant to the specific task at the receiver. Motivated by the recent advancements of deep learning, semantic communications achieve joint source channel coding through end-to-end deep neural networks, proving to deliver superior performance for diverse tasks and overcoming the ``cliff effects'' and the ``leveling effects'' of traditional separate source coding and channel coding \cite{bourtsoulatze2019deep}. Many existing works have investigated the potential of semantic communications for supporting text transmission\cite{xie2020lite,xie2021deep,getu2024performance}, image transmission,\cite{bourtsoulatze2019deep,xu2021wireless,yang2023witt}, and even channel state information \cite{xu2022deep}, enabling data compression while reserving the effectiveness. Some studies also investigated the relationships between multi-modal data to facilitate multi-modal or multi-task semantic communications \cite{xie2022task,zhang2022unified,zhu2024multi}. As a crucial data format, point clouds have been the focus of several studies on semantic communications. A wireless point cloud delivery method using graph neural networks was proposed in \cite{fujihashi2021wireless}. The authors in \cite{liu2023semantic} developed a point cloud semantic communications system with controllable coding rate for multi-user transmission. The point-transformer-based network was proposed to transmit point clouds over wireless channels with limited bandwidths \cite{bian2023wireless}. However, these studies focus on point clouds of regular objects, such as those in ShapeNet \cite{chang2015shapenet} and ModelNet40 \cite{wu20153d}, and are not directly applicable to LiDAR point cloud transmission.

In this paper, we aim to design a digital transmission system for LiDAR point cloud, which differentiates itself from most existing works on semantic communications. We develop an end-to-end LiDAR point cloud transmission network incorporating feature enhancement and feature fusion. Furthermore, we assume digital transmission with quantization and modulation, and adopt a simple yet effective nonlinear activation layer to cope with digital communication noise, distinguishing our approach from DTAT widely used in semantic communications.

\section{System Model}
We consider transmitting LiDAR point clouds of the traffic scenes over the physical channel between the agents equipped with LiDAR sensors. For simplicity, we asssume there are two connnected agents, one acting as the transmitter and the other as the receiver. The input of the system is the LiDAR point cloud generated by the transmitter agent, $\mathbf{P}^T=[\mathbf{p}_1, \mathbf{p}_2, ..., \mathbf{p}_N]$, where $N$ is the number of points of a point cloud and $\mathbf{p}_n$ is a reflected point of LiDAR, $\mathbf{p}_n\in \mathbb{R}^{3}$. The number of points $N$ depends on the LiDAR hardware configuration and is usually very large. To handle the huge amount of data, the transmitter maps the LiDAR point cloud into an effective, representative and robust feature, which is then transmitted through the physical wireless channel. The noisy received feature is decoded at the receiver and fused with the feature extracted from the receiver's point cloud $\mathbf{P}_R$ to enhance reconstruction quality. The fused feature is further processed to recover the transmitter point cloud $\mathbf{P}_T$, thereby improving the perception capability of the receiver.

Particularly, the transmitter consists of two components, feature encoder and channel encoder. The feature encoder compresses the transmitter point cloud into a compact and effective representation. According to existing deep learning-based point cloud compression methods, the compressed representation can be written as $(\mathbf{P}_s, \mathbf{F}_s)$, where $\mathbf{P}_s\in\mathbb{R}^{N_s\times 3}$ represents coordinates, and $\mathbf{F}_s\in\mathbb{R}^{N_s\times C}$ represents features of the downsampled point set. $N_s$ is the number of points of the downsampled points depending on the number of downsample layers and the downsample ratio. We assume the transmission of coordinates is error-free, because we can define an \textit{anchor point set} that the transmitter and the receiver have already obtained before the transmission. The channel encoder aims to exploit the cross-point correlations among the features and guarantees successful transmission of LiDAR point cloud features over the physical channel. The encoded feature can be represented by 
\begin{equation}
\mathbf{x}=\text{ENC}(\mathbf{P}_T)=\mathcal{F}(\mathcal{C}_{\beta}(\mathcal{S}_{\alpha}(\mathbf{P}_T)))
\end{equation}
where $\mathbf{x}\in\mathbb{R}^{N_s\times C}$,  $\mathcal{S}_{\boldsymbol{\alpha}}$ is the compression network with the parameter set $\mathbf{\boldsymbol{\alpha}}$ and $\mathcal{C}_{\boldsymbol{\beta}}$ is the channel encoder with the parameter set $\mathbf{\boldsymbol{\beta}}$. The feature $\mathbf{x}$ is then quantized to discrete values and modulated to digital symbols $\mathbf{s}$, which can be represented as 
\begin{equation}
\mathbf{s}=\text{MOD}(\mathbf{x})
\end{equation}
where $\mathbf{s}\in\mathbb{C}^{N_{s}C\times 1}$, $\text{MOD}$ represents the quantization and modulation process. Then, $\mathbf{s}$ is sent through physical channel and we assume the coherent time is ${N_{s}C}$, which is also called block fading channel. The received signal is given by
\begin{equation}
\mathbf{\hat{s}}=h\mathbf{s}+\mathbf{n}
\end{equation}
where $\mathbf{\hat{s}}\in\mathbb{C}^{N_{s}C\times 1}$, $h$ represents the Rayleign fading channel with $\mathcal{CN}(0,1)$ and $\mathbf{n}\sim\mathcal{CN}(0, \sigma^2\textbf{I})$. We assume channel $h$ is known, then the received symbol can be processed by channel equalization
\begin{equation}
\widetilde{\mathbf{s}}=\frac{h^{*}}{|h|^2}\mathbf{\hat{s}}=\mathbf{s}+\frac{h^{*}}{|h|^2}\mathbf{n}
\end{equation}
where we use zero-forcing (ZF) equalization. Then through demodulation, the received feature can be represented by 
\begin{equation}
\mathbf{y}=\text{DEMOD}(\mathbf{\widetilde{s}})
\end{equation}
where $\text{DEMOD}$ represents the demodulation and dequantization process. 

For end-to-end training of the system, the modulation, the channel, and the demodulation must allow back-propagation. We use the straight-through estimator\cite{bengio2013estimating} to model the impact of digital communciation system as equivalent noise given by
\begin{equation}
\mathbf{\hat{x}}=\mathbf{x}+sg(\mathbf{y}-\mathbf{x})\label{ste}
\end{equation} 
where $sg$ represents the stop gradient operator that detaches the gradient of tensors in the computation graph. The receiver consists of three parts as well, including channel decoder, feature fusion, and feature decoder. The structure of the channel decoder is the same as the channel encoder which can exploit the valuable information from the noisy received features. The feature fusion module is to enhance the transmitted point cloud feature by fusing the feature extracted from the receiver point cloud based on the spatial correlations from different perceptive perspectives. The reconstructed process at the receiver can be represented by 
\begin{equation}
\mathbf{\hat{P}}_T=\text{DEC}(\mathbf{\hat{x}}, \mathbf{P}_R)=\mathcal{S}^{-1}_{\boldsymbol{\gamma}}(\mathcal{F}_{\boldsymbol{\theta}}(\mathcal{C}^{-1}_{\boldsymbol{\delta}}(\hat{\mathbf{x}}), \mathbf{x}_R))
\end{equation}
\begin{equation}
\mathbf{x}_R = \mathcal{S}_{\boldsymbol{\alpha}}({\mathbf{P}_R})
\end{equation}
where $\mathbf{\hat{P}}_T$ represents the reconstructed point cloud, $\mathbf{x}_R$ is the feature extracted from the receiver point cloud using the same feature encoder as the transmitter, $\mathcal{C}^{-1}_{\boldsymbol{\delta}}$ is the channel decoder with the parameter set $\boldsymbol{\delta}$, $\mathcal{F}_{\boldsymbol{\theta}}$ is the feature fusion module with the parameter set $\boldsymbol{\theta}$ and $\mathcal{S}^{-1}_{\boldsymbol{\gamma}}$ is the feature decoder with the parameter set $\boldsymbol{\gamma}$. 

The goal of the system is to minimize the difference between the original point cloud and the reconstructed point cloud given the constraint of the fixed number of transmitted symbols under different wireless communication conditions. We utilize the point-to-point Chamfer Distance to measure the reconstruction error. Due to the complex structure of LiDAR point clouds, only using Chamfer Distance cannot capture the structure of point clouds very well, so we apply a density-preserving loss function\cite{he2022density} that will be demonstrated in detail in the next section.

In summary, according to the problem formulation, we need to address three key challenges for the design of LiDAR point cloud feature transmission system: i) design the LiDAR point cloud feature encoder (decoder) network $\mathcal{S}_{\alpha}$ to achieve high compression ratio, ii) design the LiDAR point cloud channel encoder network $\mathcal{C}_{\beta}$ and feature fusion network $\mathcal{F}_{\theta}$ to  enhance reconstruction quality against wireless channel noise, and iii) mitigate the effects of non-differentiable operations in digital communications. The core objective of this paper is to tackle these challegnes and design an efficient LiDAR point cloud feature transmssion system to support time-sensitive multi-agent collaborative perception.

\section{LiDAR Point Cloud Feature Transmission for Collaborative Perception}
In this section, we introduce our proposed deep learning-enabled communication system for LiDAR point cloud transmission, that is, LiDAR Point Cloud Feature Transmission, termed LPC-FT. The proposed LPC-FT is shown in Fig. \ref{img:overview}. We adopt the state-of-the-art density-preserving point cloud compression neural network to extract task-aware features. In order to overcome the impariment caused by the wireless channel, we propose the channel encoder network and feature fusion network inspired by the attention mechanism for point clouds. Besides, we utilize the nonlinear activation layer to fit the digital communication system, and apply transfer learning to improve the training efficiency across different channel conditions.

\subsection{Feature Encoder and Decoder}
\label{subsection: semantic}
\begin{figure*}
    \centering                                   
    \includegraphics[width=1.0\textwidth]{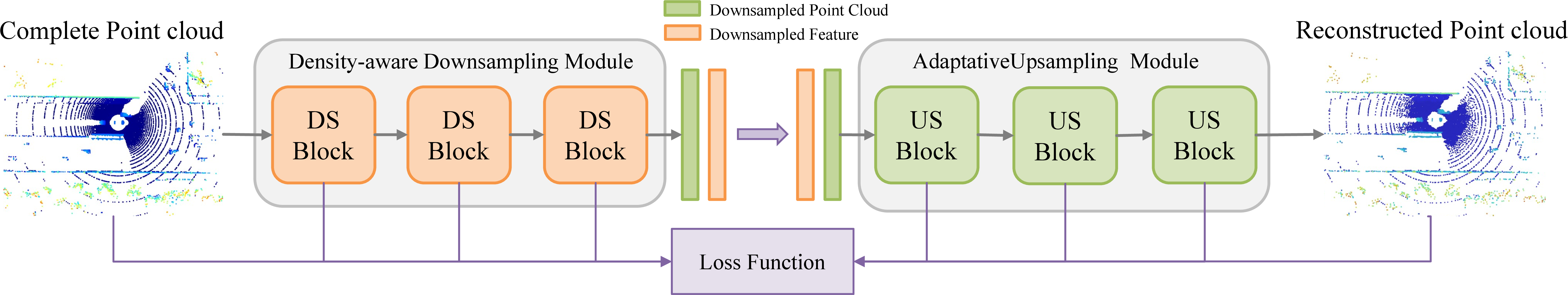}    
    \caption{The architecture of the feature encoder and decoder, which consists of downsampling blocks and upsampling blocks.}
    \label{img:semantic_encoder} 
    \vspace{-0.5cm}
\end{figure*}

Given the sparsity, complexity, and non-uniform distribution of LiDAR point clouds, it is necessary to design a feature extraction neural network that addresses these characteristics. Accordingly, we utilize a density-preserving deep point cloud compression framework as the feature encoder and decoder. This framework is based on an auto-encoder architecture, as illustrated in Fig. \ref{img:semantic_encoder}. 

The encoder consists of $n$ cascaded downsampling blocks, which can aggregate the features of the entire point cloud to a downsampled point set $\mathbf{P}_s$ and a representative feature set $\mathbf{F}_s$. At each stage of the encoder, an input LiDAR point cloud will be downsampled by a factor of $f_s$ using farthers point sampling (FPS), which is applied in point cloud processing widely. In order to address the problem of reconstruction with poor accuracy and uniform density, each downsampling block caculates three different embeddings: density embdding, local position embedding and ancestor embedding. The density embedding module caculates the cardinality of the neighbourhood point set and maps the cardinality to an embedding via MLPs. The local position embedding is to capture the distribution of the neighbourhood point set which encodes the directions and distances of these points to an embedding based on attention mechanism. The ancestor embedding module utilizes the point transformer to aggregate the features of the neighbourhood point set from the previous stage to the representative downsampled point. The point transformer is based on vector self-attention. For the downsampled point $\mathbf{p}_i$, the point transformer can be represented as:

\begin{equation}
\mathbf{y}_i=\sum_{\mathbf{p}_j\in \mathbf{P}(\mathbf{p}_i)}Softmax(\gamma(\phi(\mathbf{f}_j)-\psi(\mathbf{f}_i)+\delta))\odot(\alpha(\mathbf{f}_j)+\delta)
\end{equation}
where $\mathbf{P}(\mathbf{p}_i)$ represents the local neighbourhood point set of $\mathbf{p}_i$, $\gamma, \phi, \psi,$ and $\alpha$ are mapping functions (e.g, MLP), $\odot$ represents the Hadamard product or the element-wise product, $\delta$ is the position encoding. The position encoding plays a vital role in the point transformer, allowing the operator to adapt to local structure in the point cloud. The vanilla position encoding can be represented as:

\begin{equation}
\delta=\theta(\mathbf{p}_i - \mathbf{p}_j)
\end{equation}
where encoding function $\theta$ is an MLP. Notably, the vanilla position encoding only encodes the relative position information. We think that it is meaningful to introduce the absolute coordinate information for the large-scale LiDAR point cloud. Thus we add the additional absolute coordinate position encoding as the following,
\begin{equation}
\delta=\theta(\mathbf{p}_i - \mathbf{p}_j)+\beta(\mathbf{p}_i)
\end{equation}
where encoding function $\beta$ is also an MLP. The absolute coordinate position encoding facilitates better feature aggregation for the ancestor embedding by considering the point position within the entire LiDAR point cloud. 
While the density and position embeddings capture local density and geometry, the ancestor embedding consolidate local features from the previous stage. Together, these embeddings can represent the multi-dimensional features of LiDAR point clouds effectively. Finally, an MLP fuses these embeddings into a new representative feature for the next downsampling stage. After $n$ downsampling stages, we can get a compact yet effective LiDAR point cloud representation, with coordinates $\mathbf{P}_s\in\mathbb{R}^{N_s\times 3}$ and features $\mathbf{F}_s\in\mathbb{R}^{N_s\times C}$.

\begin{figure}
    \centering                                   
    \includegraphics[width=0.95\linewidth]{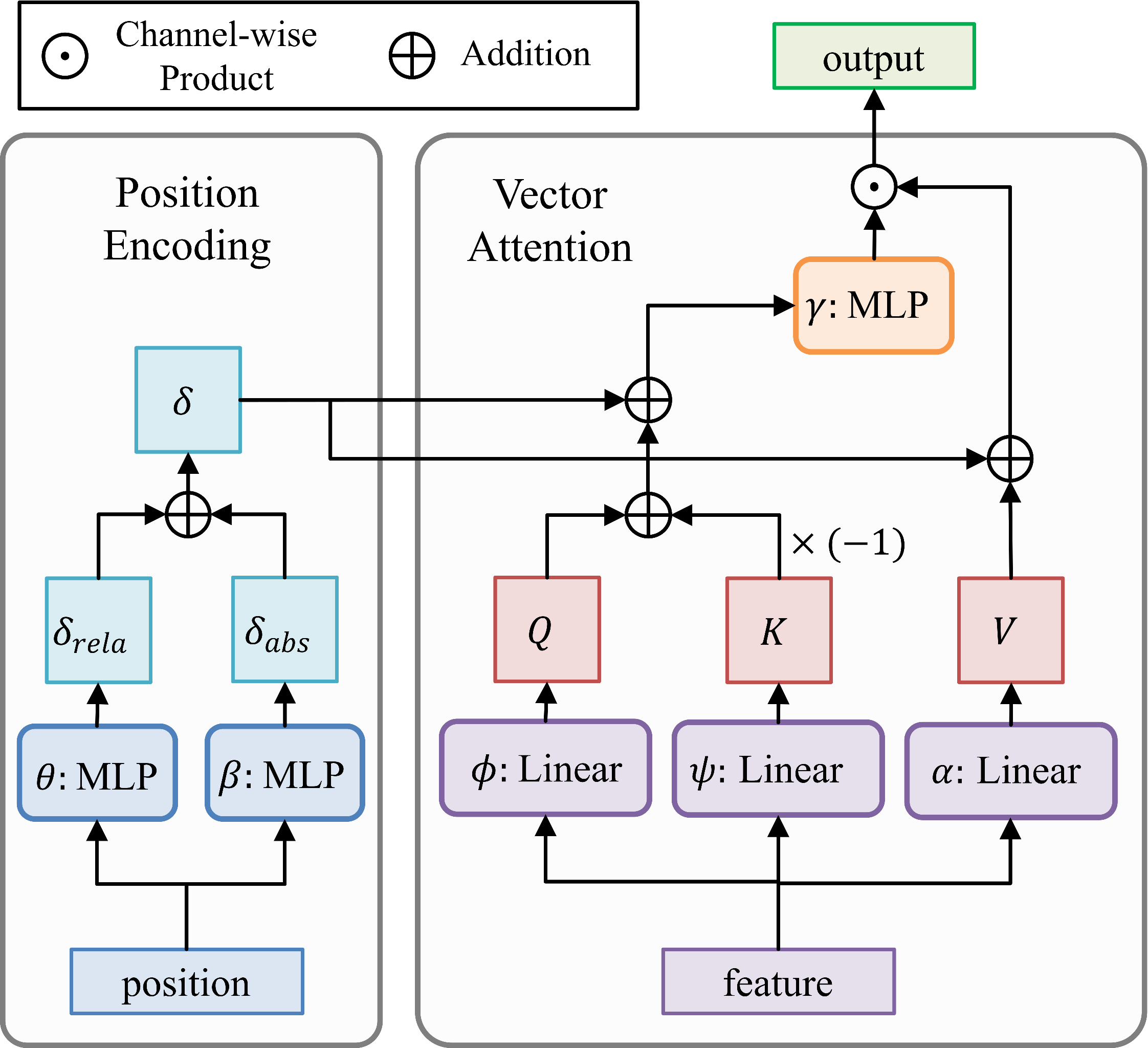}    
    \caption{The improved point transformer block with the absolute coordinate position encoding. The improved block can aggregate the point features from the previous stages more efficiently with the integration of absolute coordinates.}
    \label{img:point-transformer} 
\end{figure}

The decoder consists of $n$ density-preserving upsampling blocks corresponding to the encoder. For point cloud reconstruction, the offset upsampling block is proposed to generate local points and features surrounding the given point\cite{wiesmann2021deep}. Centering at each point $\hat{\mathbf{p}}_i\in\mathbf{P}_s$, the original block will predict $K$ offset coordinates $\Delta\mathbf{P}_s(\hat{\mathbf{p}}_i)\in\mathbb{R}^{K\times 3}$ within the given radius $r$, where $K$ is a fixed upsampling ratio and $r$ is a fixed scale factor. The final predicted points can be represented as the upsampled neighbourhood point set 
\begin{equation}
\hat{\mathbf{P}}(\hat{\mathbf{p}}_i)=\hat{\mathbf{p}}_i+\Delta\mathbf{P}_s(\hat{\mathbf{p}}(i)) \label{eq:upsampling}
\end{equation}
To enable the propogation of feature information $\mathbf{f}_i\in\mathbf{F}_s$ during the upsampling, the block will also utilize MLPs to generate the upsampled feature set $\hat{\mathbf{F}}(\hat{\mathbf{p}}_i)\in\mathbb{R}^{K\times C}$. As a result, by stacking a few upsampling blocks, the point clouds can be reconstruceted incrementally. 

However, LiDAR point clouds are highly non-uniform, that different local regions have different geometry properties. So we utilize a density-preserving upsampling block as \cite{he2022density}, whose upsampling ratio $K$ and scale factor $r$ are learnable through processing different point features. Because the encoder has aggregated the density and the local position information to the feature embedding, the decoder block has the potential to predict the customized upsamping ratio and scale factor. Furthermore, a novel operator sub-point convolution is facilitated to prevent the feature replication and reduce the computations, resulting in more efficient reconstruction. In summary, the output of each point $\hat{\mathbf{p}}_i$ at the scalable upsampling stage can be represented as 
\begin{equation}
(\hat{\mathbf{P}}(\hat{\mathbf{p}}_i), \hat{\mathbf{F}}(\hat{\mathbf{p}}_i), K_i), \hat{\mathbf{p}}_i\in\mathbf{P}_s, K_i\leq K_{max}
\end{equation}
where $K_i$ is the predicted upsampling ratio for the point $\hat{\mathbf{p}}_i$, $\hat{\mathbf{P}}(\hat{\mathbf{p}}_i)$ and $\hat{\mathbf{F}}(\hat{\mathbf{p}}_i)$ both have $K_{max}$ items, but only the first $K_i$ points and features will be selected. The union of all the selected points is the upsampled point set for the next upsampled stage and this is also suitable for the features. Through $n$ upsampling stages, we can reconstruct a density-preserving LiDAR point cloud. 

\begin{figure*}
    \centering                                   
    {\includegraphics[width=0.95\linewidth]{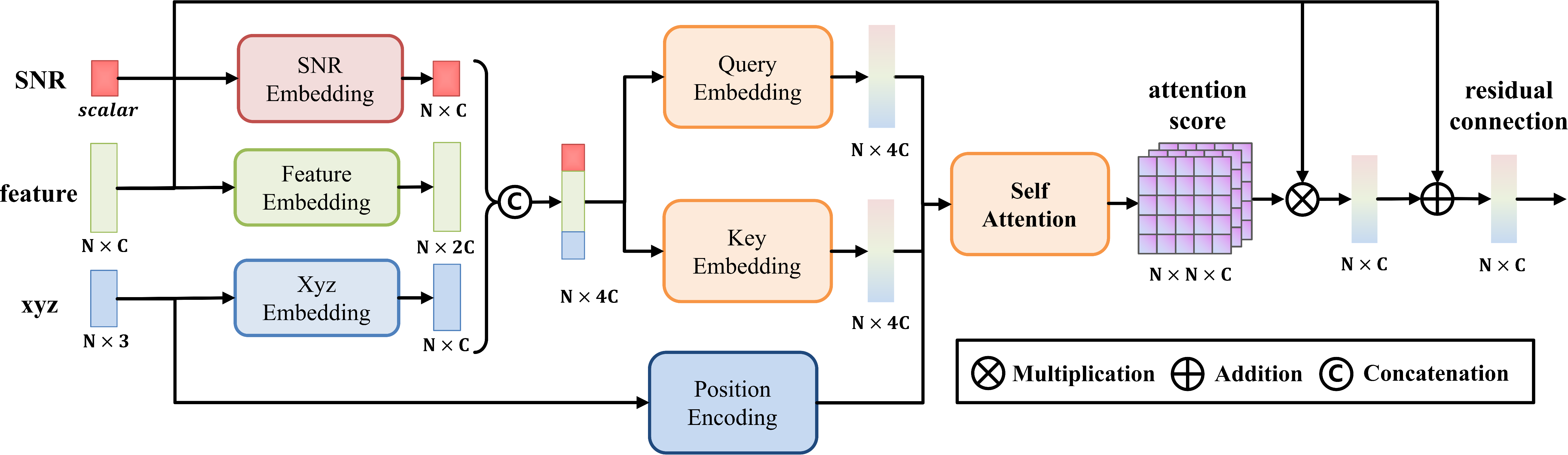}}
    \caption{The channel encoder network for the LiDAR point cloud transmission based on self-attention mechanism. Each embedding submodule is an MLP with a specific hidden layer configuration, which transforms and aligns different features.}
    \label{img:channel_encoder}
\end{figure*}

\subsection{Channel Encoder and Decoder}
\label{subsection: channel}
To reduce the effects of channel noise, we propose a novel self-attention-based channel encoder network specifically designed for LiDAR point clouds. The channel decoder shares the same architecture as the channel encoder. As described in section \ref{subsection: semantic}, the feature encoder is a point-wise network that processes each point's feature independently. Consequently, the channel encoder module aims to exploit the correlations of different points and enhance the task-aware features produced by the feature encoder. The feature enhancement process should be adaptable to two variables: point features and channel conditions. For adaptation to point features, the module should capture the varying weights between different points. For instance, points located on the same strcture, such as walls or trees, should exhibit strong correlations. For adaptation to channel conditions, the module should produce varying weights for fusion under different SNRs. For example, when SNR is high, the weight of the point set should be uniform. In such scenarios, the effects of transmission errors are minimal, and the network tends to utilize the nearly accurate features of different points. Based on these two adaptation capabilities, the channel encoder network can enhance features adaptively and remain the high performance of reconstruction against channel noise.

Fig. \ref{img:channel_encoder} shows the network architecture of the channel encoder network. The network has three inputs: point positions $\mathbf{P}_s$, point features $\mathbf{F}_s$, and SNR $\mathbf{\sigma}$. Firstly, we use MLPs to map these inputs to different feature embeddings, position embedding $\mathbf{E}^p\in\mathbb{R}^{N_s\times C}$, feature embedding $\mathbf{E}^f\in\mathbb{R}^{N_s\times 2C}$, and SNR embedding $\mathbf{E}^{\sigma}\in\mathbb{R}^{N_s\times C}$ respectively. We expect this mapping can align them in the same feature space. Then we concatenate them together as a unified embedding $\mathbf{E}^u\in\mathbb{R}^{N_s\times4C}$ 
\begin{equation}
\mathbf{E}^u=Concat({\mathbf{E}^{p}, \mathbf{E}^{f}, \mathbf{E}^{\sigma}})
\end{equation}
where $Concat[.,.]$ represents the concatenation operation along the channel dimension. We can modulate the weights of different points based on this unified representation. To exploit the correlations between different points adaptaively, we utilize the self-attention mechanism. We use MLPs for query and key computation instead of the linear projections employed in typical transformers. This substitution is made because we believe that the nonlinear operations of MLPs can derive more representative features from the concatenated inputs. Consequently, the unified feature embedding $\mathbf{E}^u$ is projected to the query embedding $\mathbf{E}^{Q}$ and the key embedding $\mathbf{E}^{K}$ through MLPs, enhancing the effectiveness of our self-attention mechanism. As illusrated in Point Transformer, we apply a substract attention operator and a relative position encoding to compute a normalized score matrix $\mathbf{A}\in\mathbb{R}^{N_s\times N_s \times C}$. We multiply the original point features $\mathbf{F}_s$ by the attention score matrix $\mathbf{A}$ in a point-wise fusion manner. Finally, we use a residual connection to combine original point features with the fused features, which is similar to the classical transformer block. Specifically, the channel encoding process can be represented as
\begin{equation}
\mathbf{f}^{c}_{i}=\mathbf{f}^{s}_{i} + \sum_{j=1}^{N_s}\mathbf{a}_{ij}\odot\mathbf{f}^{s}_{j}
\end{equation}
\begin{equation}
\mathbf{a}_{ij}=Softmax (\mathbf{e}_{i}^{Q}-\mathbf{e}_{j}^{K}+\xi (\mathbf{p}^{s}_i - \mathbf{p}^{s}_{j}))
\end{equation}
where $\mathbf{f}^{c}_{i}\in \mathbb{R}^{C}$ represents the output feature of the point $\mathbf{p}_i^s$, $\mathbf{f}_{i}^{s}\in \mathbb{R}^{C}$ represents the original feature of the point $\mathbf{p}_i^s$ and the same goes for $\mathbf{f}_{j}^{s}$, $\mathbf{a}_{ij}\in\mathbb{R}^{C}$ represents the channel-wise attention score between $\mathbf{p}_i^s$ and $\mathbf{p}_j^s$ in the self-attention matrix $\mathbf{A}$.

\begin{figure*}
    \centering
    {\includegraphics[width=0.95\linewidth]{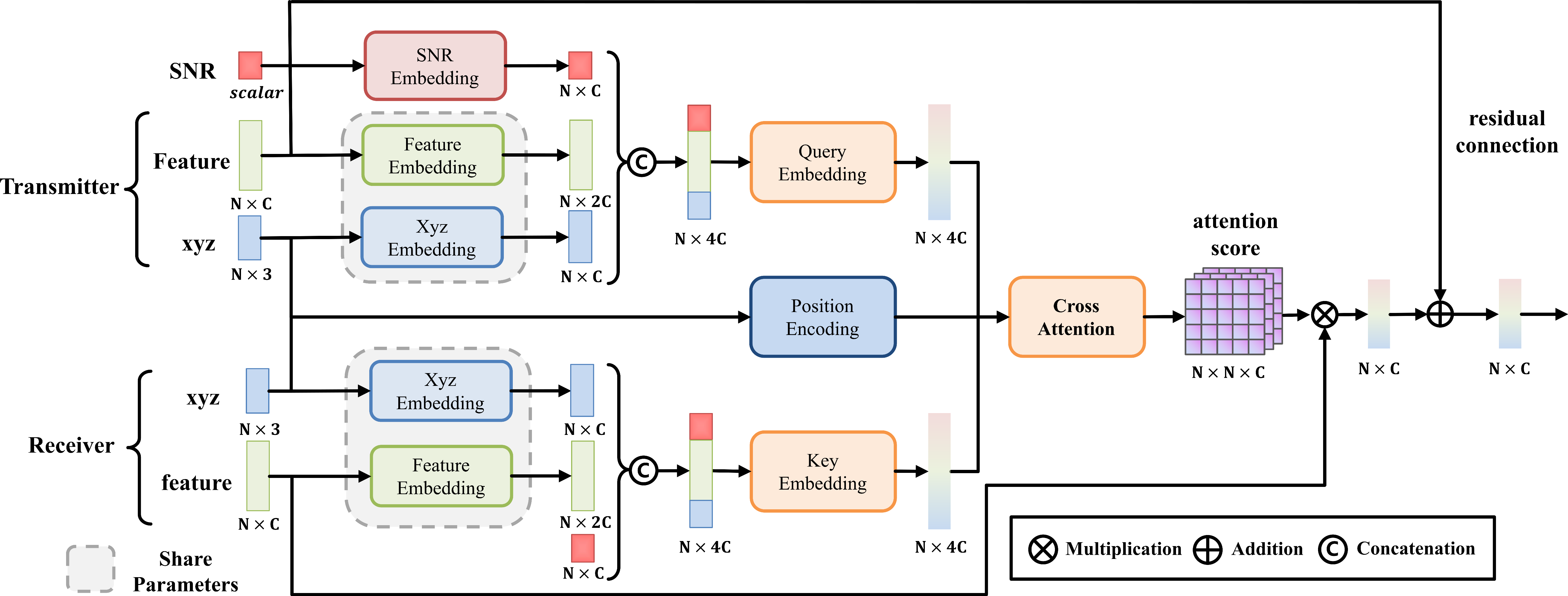}}  
    \caption{The feature fusion network for the transmitter and receiver based on cross-attention mechanism. Each embedding submodule is an MLP with a specific hidden layer configuration similar to the channel encoder network. The transmitter and receiver share the network parameters for feature preprocessing as decipted by the grey box.}
    \label{img:feature fusion}   
\end{figure*}

\subsection{Feature Fusion Module}
\label{subsection: fusion}
For multi-agent collaborative perception tasks, there are diverse sensors deployed on the robots, vehicles, or road side units (RSU). In order to exploit the abundant environment information from multiple agents, we propose a feature fusion module deployed on the receiver to fuse the point cloud features from the receiver and the transmitter. The transmitter and the receiver perceive the environment from different perspectives, and their point clouds might describe the same structure or object in some regions. Therefore, we can utilize the features extracted from the receiver point cloud to improve the reconstruction quality of the transmitter point cloud. Numerous studies in collaborative perception have concentrated on the fusion of bird's-eye view (BEV) features \cite{li2021learning,xu2022v2x,wang2023core}. These approaches involve aligning the BEV feature maps from different agents and performing operations on the features within the same grid. In contrast, our compression network is a point-wise network that directly manipulates the coordinates of unordered points. Because the points from different agents lack clear correspondence, the operations utilized in BEV feature fusion methods are not directly applicable. To overcome this issue, we propose a point-wise feature fusion network that leverages the cross-attention mechanism. The cross-attention operation computes an $N\times N$ attention score matrix between the transmitter and the receiver, automatically exploiting relationships without the need for point-to-point correspondence.

The architecture of the feature fusion network is depicted by Fig. \ref{img:feature fusion}. The network has five inputs, that is, the channel decoder output feature of the transmiter point cloud $\mathbf{F}^{T}$ and the corresponding point set $\mathbf{P}^{T}$, the feature encoder output feature of the receiver point cloud $\mathbf{F}^R$ and the corresponding point set $\mathbf{P}^{R}$, and the channel SNR $\sigma$. Similar to the channel encoder, we first use MLPs to project these inputs to their corresponding feature embeddings. Additionally, we apply the same network parameters for both the transmitter and receiver feature projections to align them within the same embedding space. We then concatenate the transmitter embeddings to obtain the transmitter unified feature embedding $\mathbf{E}^{T_u}$. Similarly, the receiver unified feature embedding $\mathbf{E}^{R_u}$ is obtained using the same method. The transmitter unified feature embedding $\mathbf{E}^{T_u}$ is projected to the query embedding $\mathbf{E}^Q$, while the receiver unified feature embedding $\mathbf{E}^{R_u}$ is projected to the key embedding $\mathbf{E}^K$. Through the cross-attention mechanism between the transmitter query embedding and the receiver key embedding, we aim to adaptatively exploit the correlations between the transceiver point clouds. The cross-attention matrix $\mathbf{A}$ instructs the module to fuse the receiver feature related to the transmitter points, there by enhancing the transmitter feature representation. Finally, we obtain the fusion feature $\mathbf{F}^f$ through the residual connection of the transmitter feature $\mathbf{F}^T$. In summary, the feature fusion process can be represented as
\begin{equation}
\mathbf{f}_i^{f}=\mathbf{f}_i^{t}+\sum_{j=1}^{N_s}\mathbf{a}_{ij}\odot\mathbf{f}_j^r
\end{equation}
\begin{equation}
\mathbf{a}_{ij}=Softmax(\mathbf{e}_i^Q-\mathbf{e}_i^K+\xi (\mathbf{p}_i^t-\mathbf{p}_j^r))
\end{equation}
where $\mathbf{f}_i^f\in\mathbb{R}^C$ represents the feature fusion output feature of the point $\mathbf{p}_i^t$, $\mathbf{f}_i^t\in\mathbb{R}^C$ represents the channel decoder output feature of the point $\mathbf{p}_i^t$ at the transmitter, $\mathbf{f}_j^r$ represents the feature encoder output feature of the point $\mathbf{p}_j^r$ at the receiver, $\mathbf{a}_{ij}\in\mathbb{R}^C$ represents the channel-wise cross-attention score between $\mathbf{p}_i^t$ and $\mathbf{p}_j^r$ in the cross-attention matrix $\mathbf{A}$.

\subsection{Nonlinear Activation Module}
\label{nonlinear}
\begin{figure}
    \centering                                   
    \subfloat[The identity straight-through estimator]{\includegraphics[width=0.95\linewidth]{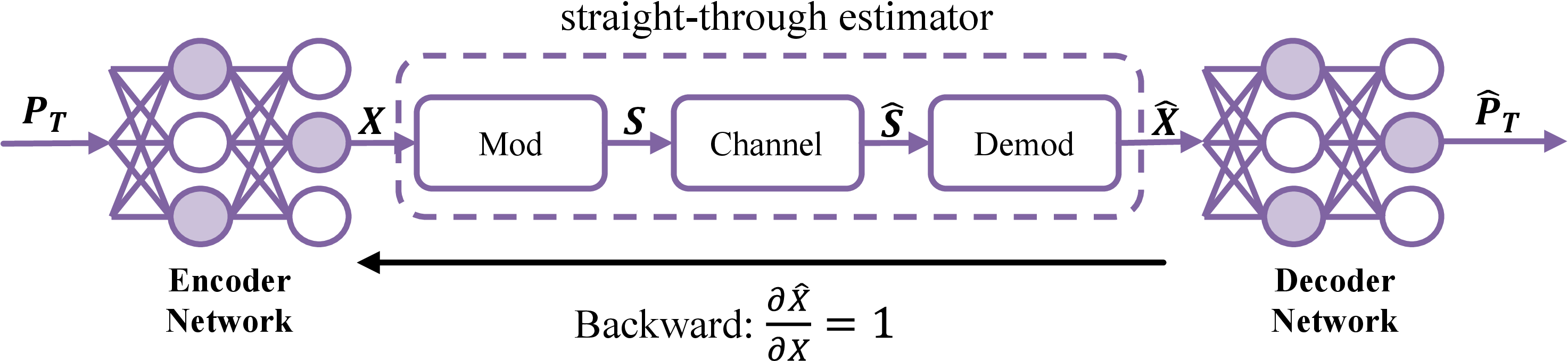}} \\
    \subfloat[The nonlinear straight-through estimator]{\includegraphics[width=1\linewidth]{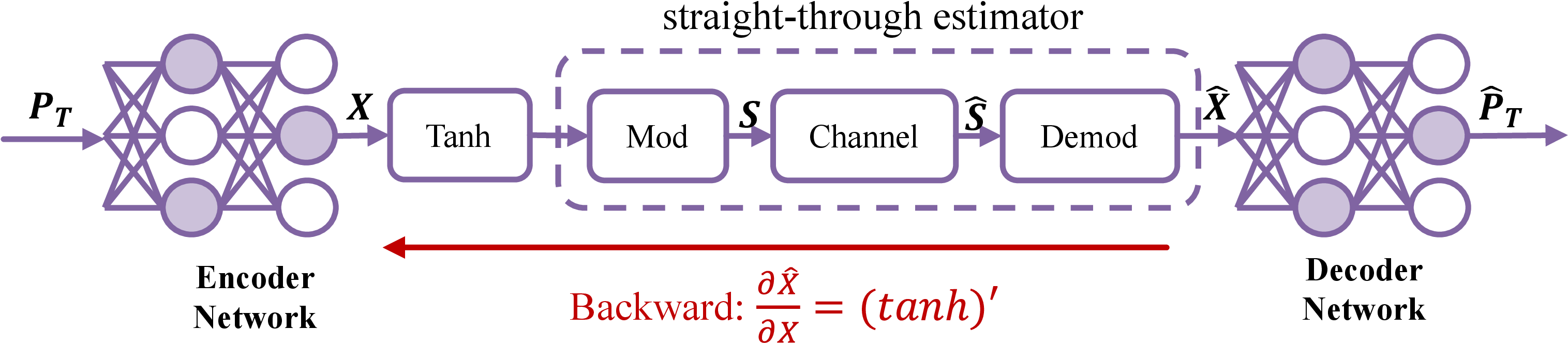}}    
    \caption{The digital communication system with the straight-through estimator}
    \label{img:ste} 
\end{figure}

Current semantic communication systems mainly focus on the discrete-time analog transmission, which omits the quantization and modualtion process present in actual digital communications. We consider a digital communication framework for LiDAR point cloud feature transmission. Digital channel noise is more challenging than analog continous noise, because a single bit error can result in significantly larger value errors. Additionally, quantization and modulation are non-differentiable processes because they operate in discrete space. The straight-through estimator is commonly used to address the non-differentiable problem in neural network training \cite{bengio2013estimating}. We employ the STE to directly copy the gradients from the received features to the transmitted features as described in equation (\ref{ste}). Through the STE operator, we can treat the impairments of the features caused by quantization, modulation and the wireless channel as equivalent noise. However, the STE operator introduces another challenge: the network lacks knowledge of the quantization and modulation processes, which can degrade task performance. To address this, we introduce a nonlinear activation layer to better fit the digital communication system, providing two key advantages related to quantization and optimization. First, improvements in nonlinear activation functions, such as $\rm{tanh}$, $\rm{sigmoid}$, have been demonstrated in previous works on quantized neural networks, addressing the gradient mismatch problem of STE \cite{gong2019differentiable}. Second, Yin \emph{et al.} \cite{yin2019understanding} proved that the nonlinear STEs can avoid the instability and incompatibility of optimization that occur with identity STE when facing binary quantized activation functions. Therefore, we add a nonlinear activation layer following the channel encoder network to generate a better feature distribution compatible with the digital communication system. Specifically, we utilize the $\rm{tanh}$ function as a nonlinear activation function to ensure the performance under digital channel noise. Fig. \ref{img:ste} illustrates the mechanism of STE with the $\rm{tanh}$ function in digital communication systems.

\subsection{Training Strategy}
\label{sec:training}
We apply a density-preserving loss function as \cite{he2022density}, which consists of three terms: the reconstruction term, the density term, and the cardinality term as follows
\begin{equation}
L=L_{cha}+\alpha L_{den}+\beta L_{card}
\end{equation}
where $\alpha$ and $\beta$ are the weights of respective terms. The reconstruction term $L_{cha}$ is the point-to-point Chamfer Distance measuring the difference between the reconstructed point cloud and the ground truth. The density term $L_{den}$ is designed to encourage recovering local density, which caculates the cardinality difference and the mean distance difference of local neighbourhood point set between $\mathbf{P}(\mathbf{p}_i)$ and $\mathbf{\hat{P}}(\mathbf{\hat{p}}_i)$. The cardinality term $L_{card}$ is designed to further measure the cardinality difference between the ground truth $\mathbf{P}_s$ and reconstructed point cloud $\mathbf{\hat{P}}_s$ at each stage. The overall loss can measure the similarity between point clouds from multiple dimensions, including distance error and local density.

We also develop a two-stage training strategy to train different components of the network, which can effectively reduce training costs. The first stage involves training the feature encoder and decoder network to compress the LiDAR point cloud without the impairments of wireless channel. During this stage, the feature encoder network acquires knowledge about the LiDAR point cloud. The second stage involves training the entire network based on the pretrained compression network from the first stage, to combat wireless channel noise and minimize the reconstruction error. When the channel condition changes, we can finetune the end-to-end transmission network starting from the pretrained feature encoder network instead of training from scratch. This approach significantly accelerates network convergence and improves the network performance.

\section{Experiment and Simulation Results}
In this section, we compare the proposed LPC-FT with other learning-based methods as well as traditional separate source coding and channel coding methods under both Additive White Gaussian Noise (AWGN) channels and Rayleigh fading channels, assuming perfect channel state informaiton (CSI) for Rayleigh fading channels. Additionally, we conduct ablation studies to verify and demonstrate the functions of the channel encoder and feature fusion module. Furthermore, we analyze the performance gains achieved through the nonlinear activation layer and transfer learning. 

\subsection{Simulation Settings}
We test our proposed method with LiDAR point clouds in OpV2V \cite{xu2022opv2v}, which is a vehicle-to-vehicle (V2V) collaborative perception simulation dataset. We choose 17 scenes containing about 7000 frames as the training set and choose other 5 scenes containing about 1000 frames as the test set. We limit the point cloud with range $[-70 \mbox{m}, 70 \mbox{m}]$ in the x and y-axis, and we sample each point cloud to $N=27648$ points using the FPS algorithm to improve the training efficiency. For the LiDAR point cloud transmission of a certain vehicle, we randomly choose another vehicle at the same timestamp in a certain commmunication range as the receiver. We set the communication range as 50 meters in all the experiments. 

For the feature encoder and decoder, we utilize three downsampling blocks and three upsampling blocks, and set the bottleneck feature dimension as 8 to achieve high point cloud compression ratio. The absolute coordinate position encoding has the same network architecture as the original position encoding, which projects the absolute 3D coordinate to a high-dimensional feature through an MLP. The other network settings of the feature encoder are the same as the density-preserving point cloud compression network\cite{he2022density}. For the channel encoder and decoder network, each submodule in Fig. \ref{img:channel_encoder} is an MLP configured with a specific hidden layer setting. We set the output dimension of the feature embedding to be twice that of the SNR embedding and coordinate embedding. This is because the features extracted by the feature encoder capture more comprehensive LiDAR point cloud information compared with original coordinates or channel conditions. For the proposed feature fusion network, we adopt the same submodule configuration as the channel encoder network. Table \ref{table:seting_training} demonstrates the hyper-parameters of pretraining and finetuning. During the fine-tuning stage for different SNRs, we utilize fewer training epochs than the pretraining stage, attributed to the well-pretrained compression model. For the density-preserving loss function, we set the density coefficient $\alpha$ as 5e-4 and the cardinality coefficient $\beta$ as 5e-6 to balance the overall reconstruction performance. The batch size is set to 1 because the performance of the adaptative upsampling decoder will degrade when the batch size increases. Our model is implemented with PyTorch, and all the deep learning simulation experimenets are trained on the Nvidia GeForce RTX 4090 GPU.

\subsection{Benchmarks}
For the baselines, we adopt the deep learning-based point cloud transmission and typical methods for separate source coding and channel coding.
\begin{itemize}
\item \textbf{Learning-based methods}: We select the existing semantic point cloud transmission method and our proposed network without considering the wireless channel as two learning-based baselines.
        \begin{itemize}
                \item \textit{Semantic Point Cloud Transmission (SEPT)}\cite{bian2023wireless}: The network consists of point transformer layers and upsampling blocks. The original SEPT processes  regular shape point clouds and utilizes the max-pooling operator to extract the global feature at the bottleneck. Because LiDAR point clouds have much more points compared with regular shape point clouds, we also adopt the \textit{anchor point set} strategy in SEPT to enhance its performance.
                \item \textit{LiDAR Point Cloud Feature Compression (LPC-FC)}: Compared with the LPC-FT, the network only contains the feature encoder and decoder. We also add the nonlinear activation layer at the bottleneck to ensure the performance facing the digital channel noise.
        \end{itemize}
\item \textbf{Traditional methods}: To perform the traditional point cloud transmission, we adopt the octree-based compression as source coding and low-density parity-check (LDPC) as channel coding. Octree-based representation is a typical point cloud compression method and is widely applied in practical application. LDPC is a classical linear error-correcting code characterized by sparse bipartite graphs. We set the code rate of LDPC as 1/2 and 3/4 respectively. Because we adopt the \textit{anchor point set} strategy, we assume the first 1/8 of bits are error-free for traditional methods to conduct the fair comparison. 
\end{itemize}

\renewcommand{\arraystretch}{1.2} 
\begin{table}
\begin{center}
\caption{The Hyper-parameters for Network Training}\label{table:seting_training}
\begin{tabular}{|>{\centering\arraybackslash}m{2cm}|>{\centering\arraybackslash}m{2cm}|>{\centering\arraybackslash}m{3cm}|}
\hline
Stage & Config & Value \\ 
\hline
\multirow{7}{*}{Pretraining} & epochs & 80 \\
  \cline{2-3}
  & optimizer & Adam \\
  \cline{2-3}
  & batch size & 1 \\
  \cline{2-3}
  & learning rate & 1e-3 \\
  \cline{2-3}
  & scheduler & StepLR \\
  \cline{2-3}
  & step size & 15 \\
  \cline{2-3}
  & gamma & 0.5 \\
  \hline
\multirow{7}{*}{Finetuning} & epochs & 20 \\
  \cline{2-3}
  & optimizer & Adam \\
  \cline{2-3}
  & batch size & 1 \\
  \cline{2-3}
  & learning rate & 1e-3 \\
  \cline{2-3}
  & scheduler & StepLR \\
  \cline{2-3}
  & step size & 4 \\
  \cline{2-3}
  & gamma & 0.8 \\
 \hline

\end{tabular}
\end{center}
\end{table}

\subsection{Performance Metric}
 We adopt the bits per point (bpp) to measure the compression ratio, that the bpp of the original point cloud is 96. Following \cite{he2022density,bian2023wireless}, we select three conventional performance metrics to measure the reconstruction quality, Chamfer Distance (CD) and two peak signal-to-noise ratio (PSNR) measures, including point-to-point (D1-PSNR) and point-to-plane (D2-PSNR). Given ground truth $\mathbf{P}_T$ and reconstructed point cloud $\hat{\mathbf{P}}_T$, we first need to caculate the point-to-point mean squared error (MSE) as follows:
\begin{equation}
\begin{split}
e^{D1}_{\mathbf{P}_T, \hat{\mathbf{P}}_T}= \frac{1}{|\mathbf{P}_{T}|}\sum_{p\in\mathbf{P}_{T}}\min_{\hat{p}\in\hat{\mathbf{P}}_{T}}\|p-\hat{p}\|_{2}^{2} \\
e^{D1}_{\hat{\mathbf{P}}_T, \mathbf{P}_T}= \frac{1}{|\hat{\mathbf{P}}_{T}|}\sum_{\hat{p}\in\hat{\mathbf{P}}_{T}}\min_{p\in\mathbf{P}_{T}}\|\hat{p}-p\|_{2}^{2} 
\end{split}
\end{equation}
Based on the caculated point-to-point MSE, we can caculate the symmetric point-to-point Chamfer Distance as:
\begin{equation}
\mathrm{CD}_{\mathbf{P}_T, \hat{\mathbf{P}}_T}=e^{D1}_{\mathbf{P}_T, \hat{\mathbf{P}}_T}+e^{D2}_{\hat{\mathbf{P}}_T, \mathbf{P}_T}
\end{equation}

Correspondingly, the caculation of point-to-point D1-PSNR is as follows:
\begin{equation}
\label{eq:d1-psnr}
\mathrm{PSNR}^{D1}_{\mathbf{P}_T, \hat{\mathbf{P}}_T}=10\log_{10} \dfrac{3\sigma^2}{\max(e^{D1}_{\mathbf{P}_T, \hat{\mathbf{P}}_T}, e^{D1}_{\hat{\mathbf{P}}_T, \mathbf{P}_T)}}
\end{equation}
where $3$ in the numerator is due to the 3D coordinates of the point cloud representation, and the peak, $\sigma$, is set to $1$ in our experiments. For the point-to-plane D2-PSNR, the caculation is very similar. We first evaluate the point-to-plane MSE as: 
\begin{equation}
e^{D2}_{\mathbf{P}_T, \hat{\mathbf{P}}_T}= \frac{1}{|\mathbf{P}_{T}|}\sum_{p\in\mathbf{P}_{T}}((p-\hat{p})\cdot \hat{n})^2
\end{equation}
where $\hat{p}$ is the nearest neighbour of $p$ in the reconstructed point cloud $\hat{\mathbf{P}}_T$, and $\hat{n}$ is the normal vector of $\hat{p}$. Following the same caculation as (\ref{eq:d1-psnr}), we can caculate the $\mathrm{PSNR}^{D2}_{\mathbf{P}_T, \hat{\mathbf{P}}_T}$.
\begin{figure*} 
    \centering                                   
    \subfloat[Chamfer Distance]{\includegraphics[height=0.33\linewidth]{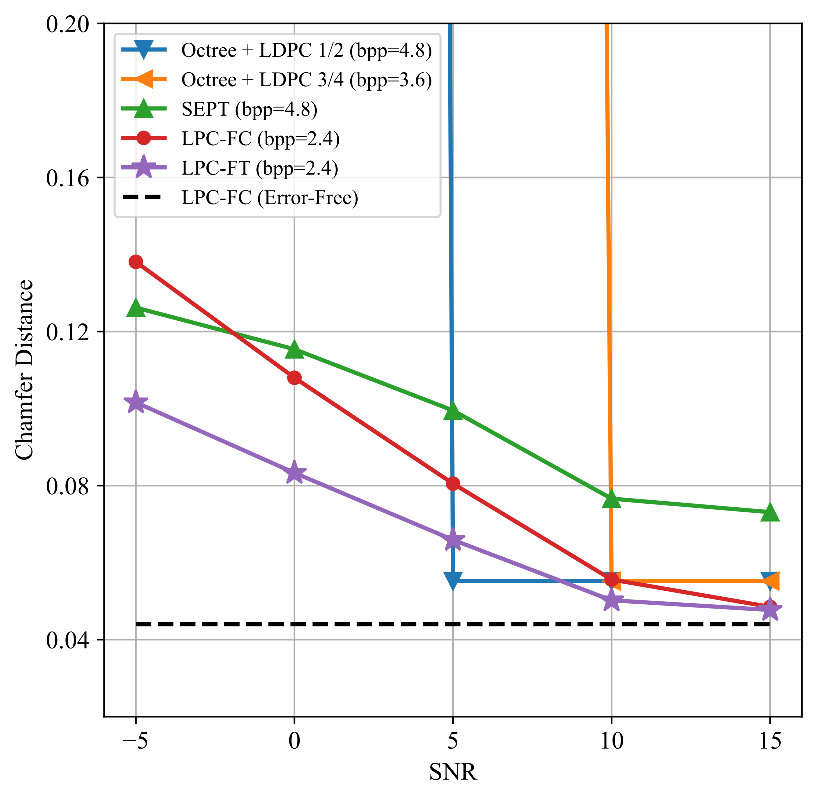}} 
    \subfloat[D1-PSNR]{\includegraphics[height=0.33\linewidth]{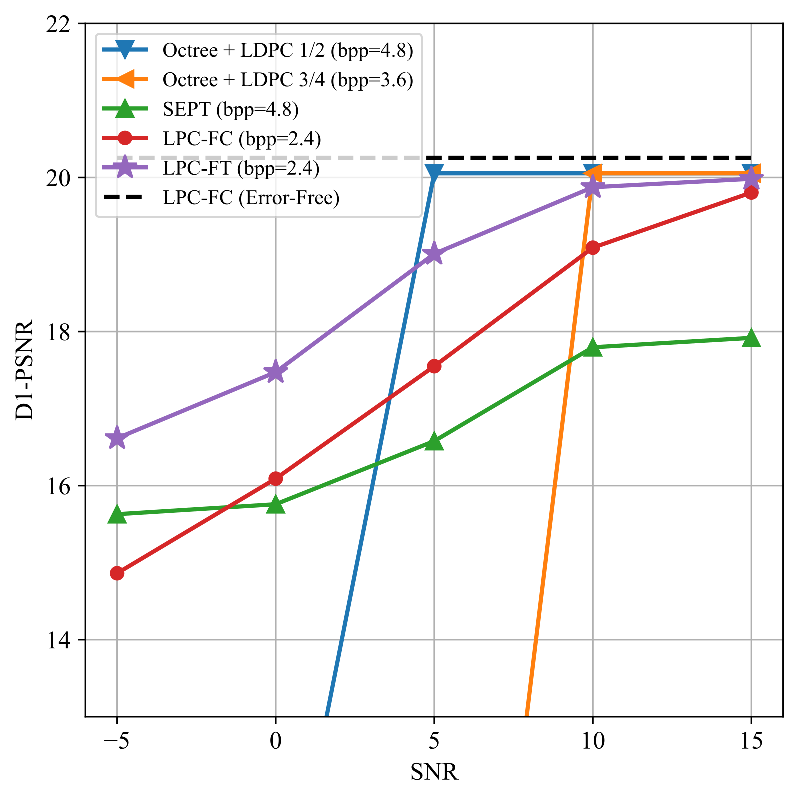}} 
    \subfloat[D2-PSNR]{\includegraphics[height=0.33\linewidth]{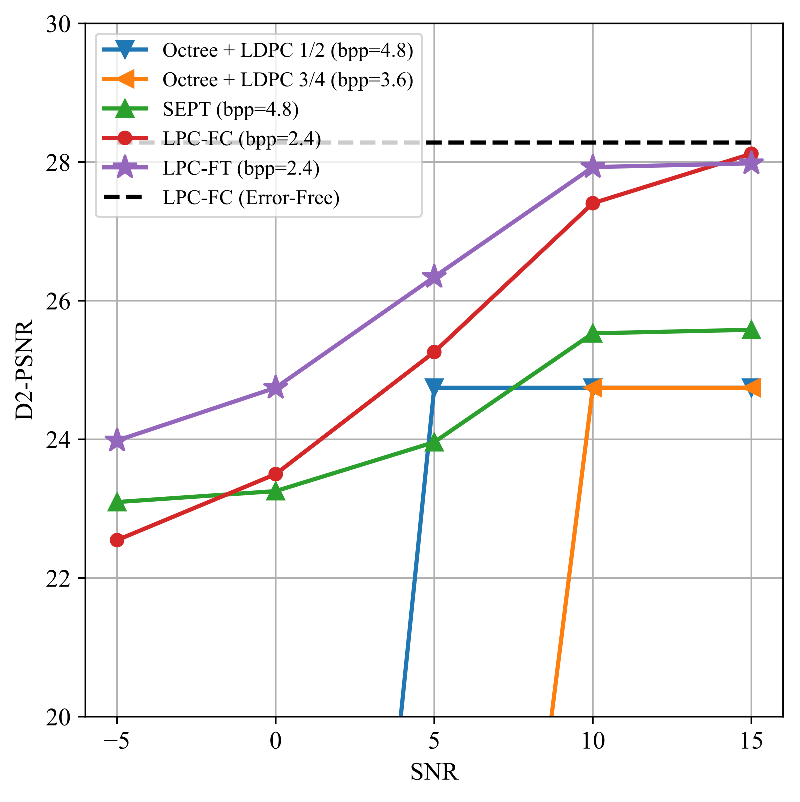}}     
    \caption{The reconstruction performance versus SNR under the AWGN channel. (a) Point-to-point Chamfer Distance. (b) Point-to-point D1-PSNR. (c) Point-to-plane D2-PSNR.}
    \label{img:performance_awgn} 
\end{figure*}
\begin{figure*} 
    \vspace{-0.25cm}
    \centering                                   
    \subfloat[Chamfer Distance]{\includegraphics[height=0.33\linewidth]{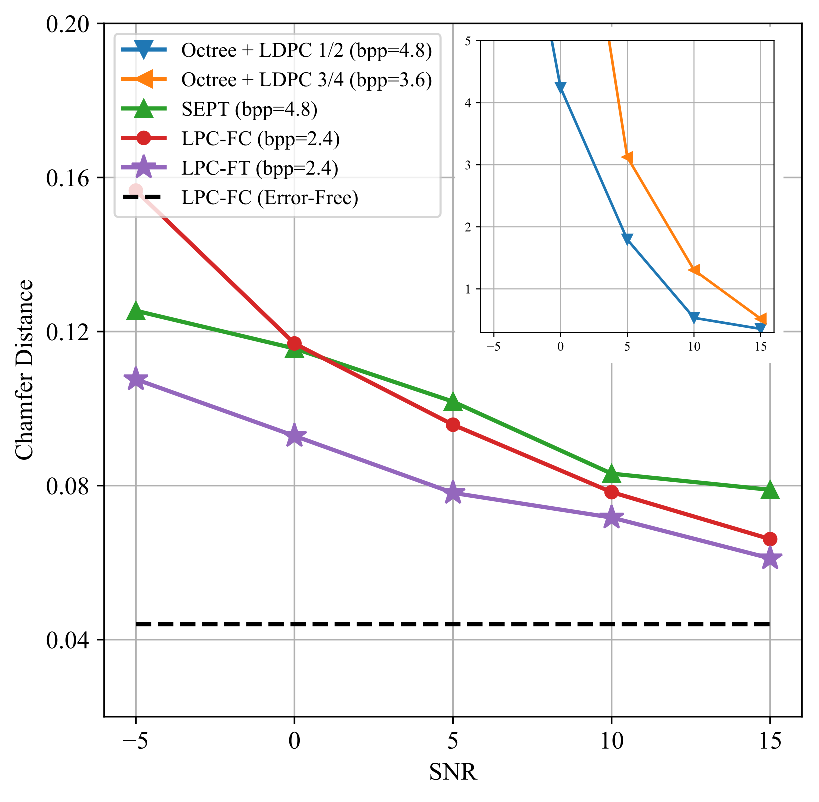}\label{img:cd_rayli}} 
    \subfloat[D1-PSNR]{\includegraphics[height=0.33\linewidth]{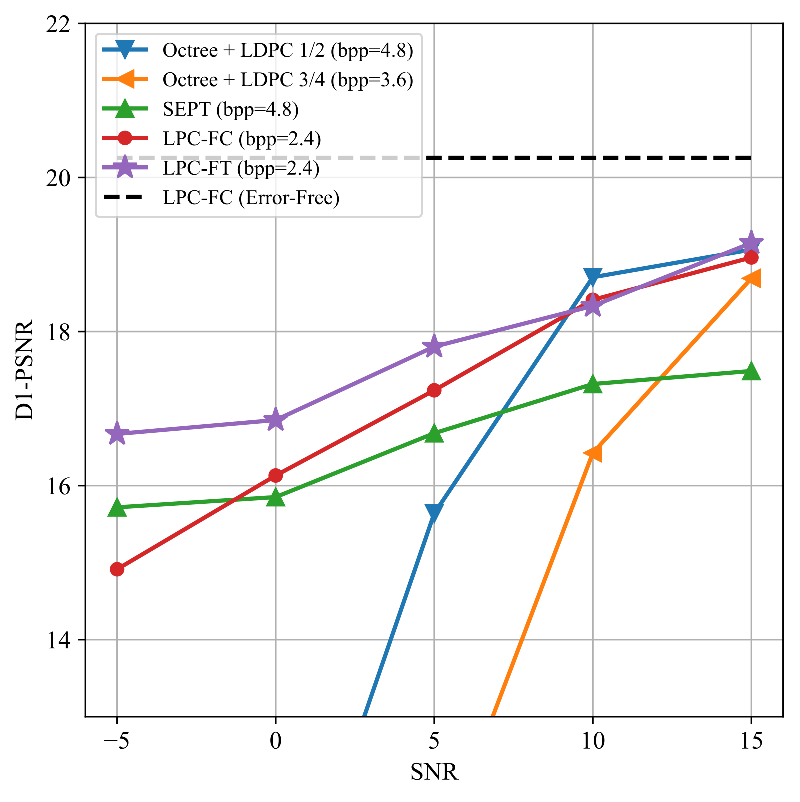}\label{img:psnr_rayli}}    
    \subfloat[D2-PSNR]{\includegraphics[height=0.33\linewidth]{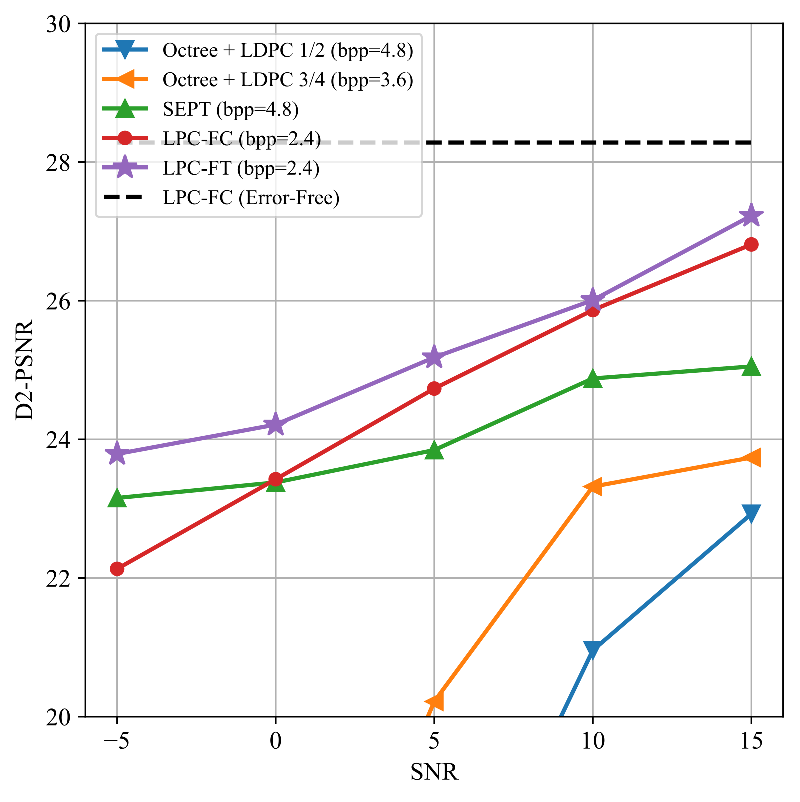}\label{img:d2psnr_rayli}}  
    \caption{The reconstruction performance versus SNR under the Rayleigh channel. (a) Point-to-point Chamfer Distance. (b) Point-to-point D1-PSNR. (c) Point-to-plane D2-PSNR.}
    \label{img:performance_rayli} 
    \vspace{-0.25cm}
\end{figure*}

\subsection{Numerical Results}
Fig. \ref{img:performance_awgn} shows the relationship between reconstruction quality and SNR under different bpps for the AWGN channel. Traditional methods achieve error-free transmission through LDPC and perform well in high-SNR regimes. The traditional method with a 1/2 code rate outperforms LPC-FT for the point-to-point metric when SNR is 5 dB, but it requires twice the transmission data compared to LPC-FT. Traditional methods also exhibit a significant performance drop when SNR falls below a certain threshold, known as the ``cliff effect''. Meanwhile, for the point-to-plane reconstruction metric D2-PSNR, traditional methods perform poorly. This is likely because the octree-based representation has low resolution for large-scale LiDAR point clouds. LPC-FT outperforms all the learning-based methods, especailly in the low-SNR regime. Compared with the SEPT, LPC-FT reduces the Chamfer Distance by an average of 30\%, improves D1-PSNR by an average of 1.9 dB, and D2-PSNR by an average of 1.9 dB, while using only half of the data. This result demonstrates the effectiveness of our proposed LiDAR point cloud transmission network, which considers the unique properties of LiDAR point clouds. We also evaluate LPC-FC to verify the efficacy of the network designed to address wireless channel noise. Compared with LPC-FC, LPC-FT reduces the chamfer distance by an average of 16\% and improves the D1-PSNR and D2-PSNR by an average of 1.1 dB and 0.8 dB, respectively. Furthermore, as SNR decreases, the performance gain becomes increasingly substantial. Compared with LPC-FC, LPC-FT achieves an SNR gain of more than 5 dB for the same reconstrcution performance in the low-SNR regime. This result demonstrates the channel coding network and the feature fusion network significantly enhance reconstruction performance against wireless channel noise. 

Fig. \ref{img:performance_rayli} shows the reconstruction performance versus SNR for Rayleigh channels. Due to the channel fading, traditional methods may fail in the face of deep fade events, even in high-SNR regimes. From Fig. \ref{img:cd_rayli}, we observe that the average Chamfer Distance of traditional methods is very large due to large reconstruction errors in deep fade events. Because PSNR is a logarithmic metric, it compresses large value ranges and reduces the impact of extreme values. Therefore, reconstruction failure cases do not significantly affect the average PSNR, as shown in Fig. \ref{img:psnr_rayli} and Fig. \ref{img:d2psnr_rayli}. Nevertheless, PSNR also decreases significantly under the Rayleigh fading channel as the SNR decreases. LPC-FT maintains high reconstruction performance among all the SNR and does not show dramatic degradation in the face of channel variations. LPC-FT also outperforms other learning-based baselines, reducing the Chamfer Distance by an average of 17\% compared with LPC-FC and 18\% compared with SEPT. In summary, Fig. \ref{img:performance_awgn} and Fig. \ref{img:performance_rayli} demonstrate that the LPC-FT achieves the best performance with the same communication data cost compared with both traditional and learning-based baselines. LPC-FT has the potential to support LiDAR point cloud transmission against wireless channel noise in multi-agent collaborative perception and other applications.

\begin{figure*} 
    \centering                                   
    \includegraphics[width=1.0\linewidth]{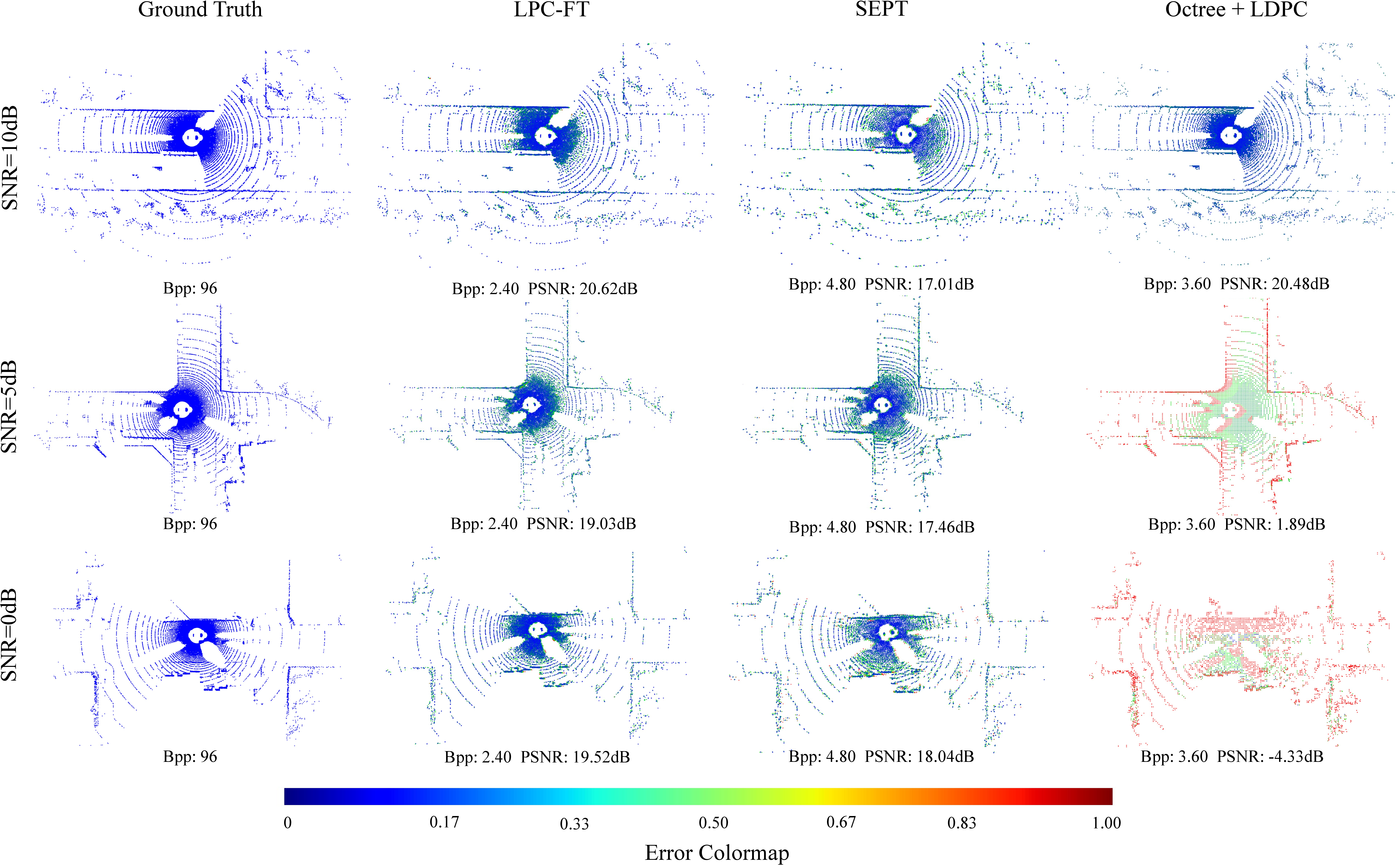} 
    \caption{Qualitative results on OpV2V. From top to bottom: SNR$=$10 dB, SNR$=$5 dB, SNR$=$0 dB. From left to right: Ground Truth, our proposed LPC-FT, existing semantic communication method SEPT\cite{bian2023wireless}, traditional method Octree-based compression with LDPC. We utilize the distance between each point in the reconstructed point cloud and its nearest neighbour point in ground truth as the error.}
    \label{img:qualitative_results} 
    \vspace{-0.25cm}
\end{figure*}

\subsection{Qualitative Results}
In this part, we analyze the qualitative results of LiDAR point cloud transmission. Fig. \ref{img:qualitative_results} shows the reconstructed LiDAR point clouds of the baseline methods and our approach for the AWGN channel. We select three point clouds from differenet scenes in OpV2V as visualization examples. The rows correspond to the reconstruction performance at different SNR levels, while the columns represent the reconstruction performance of various methods. The color of the point represents the normalized distance between the point in the reconstructed point cloud and its nearest point in ground turth. From Fig. \ref{img:qualitative_results}, we observe that the reconstructed point clouds of SEPT does not retain the local density information of the original point cloud, and exhibits severe distortion at low SNR levels. Traditional methods achieve outstanding reconstruction performance when SNR is 10 dB. Nevertheless, the reconstructed point clouds become sparse, cluttered and globally shifted when transmission errors occur. This issue arises because nodes are interdependent in an octree-based point cloud structure, and an error in one node can influence the states of other nodes. Through the task-aware network design and end-to-end network training, LPC-FT achieves favourable reconstruction results close to the ground truth, even when SNR is 0 dB. 
\begin{table}[tbp]
\begin{center}
\caption{Ablation Study of Absolute Position Encoding}\label{table:abs}
\begin{tabular}{|c|ccc|}
\hline
Metric & Chamfer Dsitance \quad & D1-PSNR \quad & D2-PSNR \quad  \\ \hline
w/ Abs. Pos. &  \textbf{0.0440} & \textbf{20.256} & \textbf{28.282}\\ 
w/o Abs. Pos. &  0.0500 & 19.539 & 27.390 \\
\hline
\end{tabular}
\end{center}
\vspace{-0.5cm}
\end{table}

\subsection{Ablation Study}
We conduct ablation experiments to analyze the roles of different components within the whole network. Table \ref{table:abs} illustrates the performance gain achieved through absolute coordinate position encoding for the LiDAR point cloud compression. We observe that incorporating absolute coordinate position encoding into the LiDAR point cloud compression network yields better performance, which reduces the Chamfer Distance by 12\%, improves the D1-PSNR by 0.7 dB, and improves the D2-PSNR by 0.9 dB. Furthermore, transmission experiments with absolute coordinate position encoding for different SNRs, as shown in the second row in Table \ref{table:ablation}, indicate that the LPC-FT achieves better reconstruction performance across various SNR levels based on the impressive compression capability.

Additionally, we validate the efficacy of the proposed channel encoder network and the feature fusion network for LiDAR point clouds through ablation experiments. As shown in Table \ref{table:ablation}, the first row represents the reconstruction perfromance of LPC-FT with both channel encoding and feature fusion. In the third row, we remove the channel encoder to analyze its performance gain. In the fourth row, we evaluate the system performance without feature fusion. The full LPC-FT achieves the best performance across all the metrics under different SNRs. We observe that at low SNR levels (SNR$=$0 dB), the performance gap between the system with only the channel encoder and the full LPC-FT is relatively small. In this scenario, wireless channel noise is the primary factor affecting reconstruction quality, and the channel encoder effectively mitigates the impact of communication noise by exploiting inter-point relationships through self-attention. Conversely, at high SNR levels (SNR$=$10 dB), the system with only feature fusion performs more closely to the full LPC-FT. When SNR is high, transmission errors are infrequent, and the channel encoder cannot significantly enhance performance through feature enhancement. However, the feature fusion module can combine the nearly accurate transmitter feature with the receiver feature without the effects of channel noise, thereby improving the reconstruction quality of LiDAR point clouds. In summary, the channel encoder plays a more significant role in the low-SNR regime through feature enhancement, while the feature fusion is more effective in the high-SNR regime by instructing the fusion of transceiver features. These results demonstrate that our proposed network design is reasonable, with different modules fulfilling their anticipated roles as described in Section \ref{subsection: channel} and \ref{subsection: fusion}. 

\begin{table*}[tbp]
\begin{center}
\caption{Ablation Study of Different Components}\label{table:ablation}
\begin{tabular}{|c|ccc|ccc|ccc|}
\hline
SNR & \multicolumn{3}{c|}{0 dB} & \multicolumn{3}{c|}{5 dB} & \multicolumn{3}{c|}{10 dB} \\ 
\hline
 Metric & CD & D1-PSNR & D2-PSNR & CD & D1-PSNR & D2-PSNR & CD & D1-PSNR & D2-PSNR \\ \hline
LPC-FT &  \textbf{0.0833} & \textbf{17.473} & \textbf{24.745} & \textbf{0.0659} & \textbf{19.010} & \textbf{26.334} & \textbf{0.0502} & \textbf{19.874} & \textbf{27.928} \\ 
w/o Abs. Pos. & 0.0862 & 17.384 & 24.667 & 0.0674 & 18.540 & 25.932 & 0.0557 & 19.191 & 27.360 \\
w/o Channel &  0.0988 & 16.676 & 23.990 & 0.0828 & 18.047 & 25.250 & 0.0566 & 19.052 & 27.483 \\ 
w/o Fusion &  0.0935 & 17.070 & 24.291 & 0.0793 & 17.785 & 25.188 &0.0632 & 18.487 & 26.642 \\
\hline
\end{tabular}
\end{center}
\vspace{-0.5cm}
\end{table*}

\subsection{Dynamic SNR Conditions}
\begin{figure} 
    \subfloat[Chamfer Distance]{\includegraphics[width=0.5\linewidth]{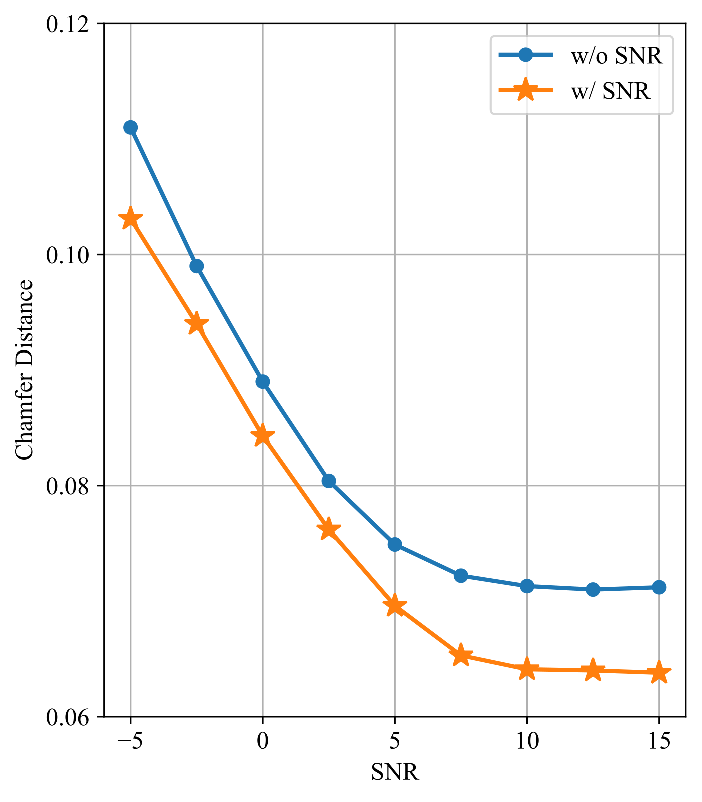}} 
    \subfloat[D2-PSNR]{\includegraphics[width=0.49\linewidth]{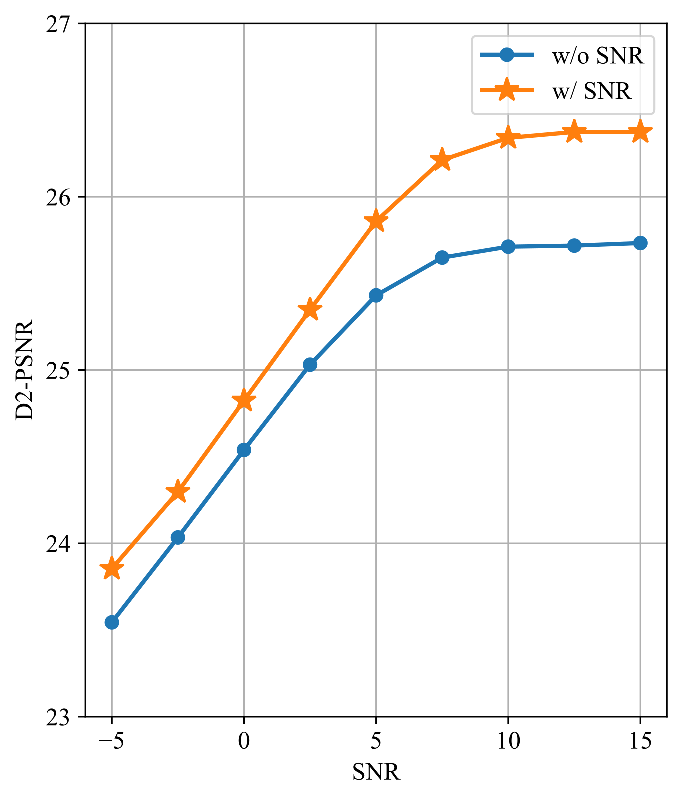}}
    \caption{The reconstruction performance under various channel conditions. (a) Point-to-point Chamfer Distance. (b) Point-to-plane D2-PSNR.}
    \label{img:cd_snr}
    \vspace{-0.25cm} 
\end{figure}

To further demonstrate the effectiveness of the SNR embedding, we conduct the experiments under under various SNR values. Specifically, we train a universal LiDAR point cloud tranmission model with a variable channel state (AWGN channel, SNR$\in$[-5,15] dB). The simulation settings are all the same except for the SNR embedding module in the channel encoder and feature fusion network. Fig. \ref{img:cd_snr} shows Chamfer Distance and D2-PSNR performance of the SNR adptative model. The SNR embedding module can reduce Chamfer Distance by an average of 7\% and improve D2-PSNR by an average of 0.45 dB. The performance gain is more prominent when SNR is high (SNR$\geq$10 dB), which can achieve a reduction of 10\% for Chamfer Distance and an improvement of 0.65 dB for D2-PSNR. Moreover, the SNR embedding module does not bring much extra computation burden with the benefit of the attention-based network design. For the channel encoder network, the SNR embedding module only brings 0.01GFLOPs improvement (from 5.62GFLOPs to 5.63GFLOPs). In summary, the results under various SNR conditions further verify the effectiveness of our proposed channel encoder network and feature fusion network for LiDAR point cloud transmission.

\begin{figure} 
    \centering                                   
    \includegraphics[width=1.0\linewidth]{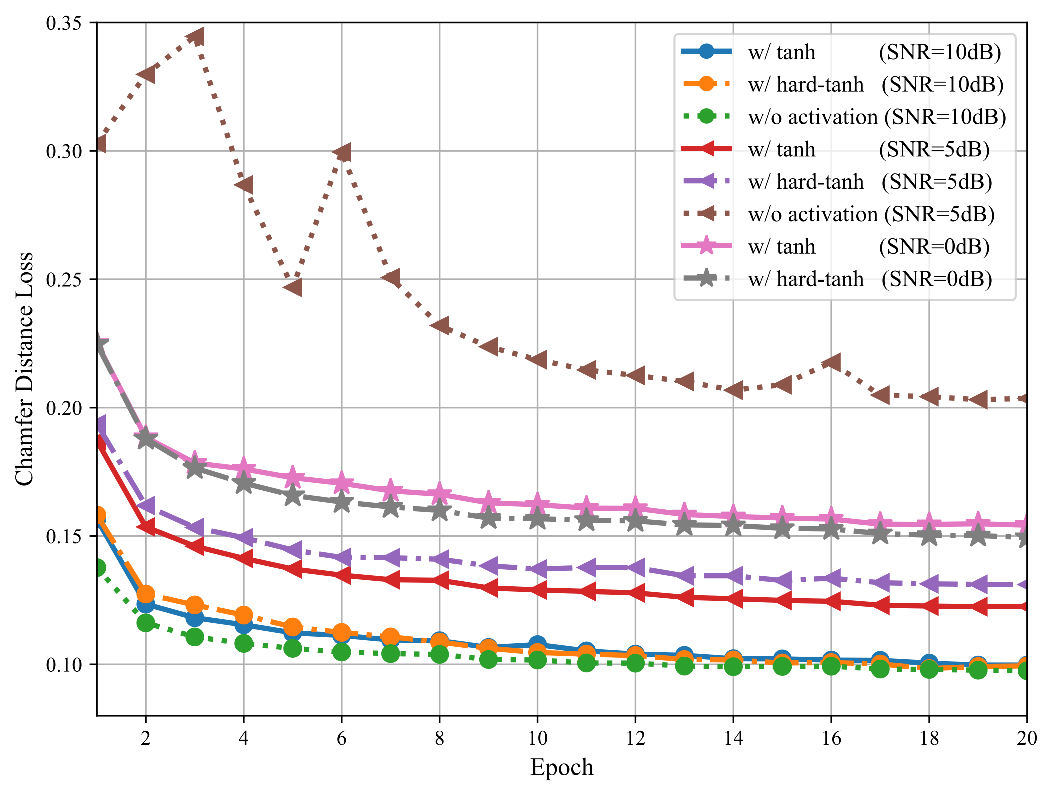} 
    \caption{Chamfer Distance loss versus training epochs for the nonlinear activation layer. The experiments are all conducted under AWGN channel.}
    \label{img:nonlinear} 
    \vspace{-0.25cm}
\end{figure}

\subsection{The Nonlinear Activation Layer}
In this section, we analyze the performance improvement provided by the nonlinear activation layer in mitigating wireless channel noise. As described in \ref{nonlinear}, we think that the nonlinear activation layer can address the gradient mismatch problem and stabilize the optimization during training for digital communication systems. To validate this, we compare our proposed network with the same network without the nonlinear activation layer. Additionally, we select another nonlinear function, $\rm{hardtanh}$, as the candidate activation layer, which can be expressed as
\begin{equation}
\mathrm{hardtanh}(x)=\begin{cases}1&\text{if}\ x>1\\-1&\text{if}\ x<\ -1\\x&\text{otherwise}\end{cases}
\end{equation}

Fig. \ref{img:nonlinear} shows the Chamfer Distance loss versus training epochs of different nonlinear layer settings under the AWGN channel. Because the network without the nonlinear activation layer fails to converge under SNR of 0 dB, its loss curve isn't shown in Fig. \ref{img:nonlinear}. When SNR is 5 dB, the loss of the network without the nonlinear activation layer remains consistently high during the training process. Additionaly, we observe the loss fluctuations that align with the optimization instability phenomenon described in \cite{yin2019understanding}. In contrast, the network with the nonlinear activation layer exhibits better training dynamics, with the training loss consistently decreasing as the epochs progress. We also observe that the performance of $\rm{tanh}$ and $\rm{hardtanh}$ is similar, indicating that the appropriate introduction of nonlinearity can significantly improve network training under digital communication conditions. Nevertheless, when SNR is 10 dB, the network without the nonlinear activation layer achieves the lowest training loss during the training process. We hypothesize that in this scenario, wireless channel noise has little impact on the network training. Therefore, the introduction of additional nonlinear activation layers may limit the feature representation capability of the network, leading to a slight decrease in reconstruction performance. In a nutshell, we believe that the nonlinear activation layer is beneficial and applicable for digital communication systems. Future research could explore designing trainable and flexible nonlinear activation layers based on the characteristics of wireless communication systems, potentially further improving task performance and system robustness against diverse communication conditions.

\begin{figure} 
    \centering                                   
    \includegraphics[width=1.0\linewidth]{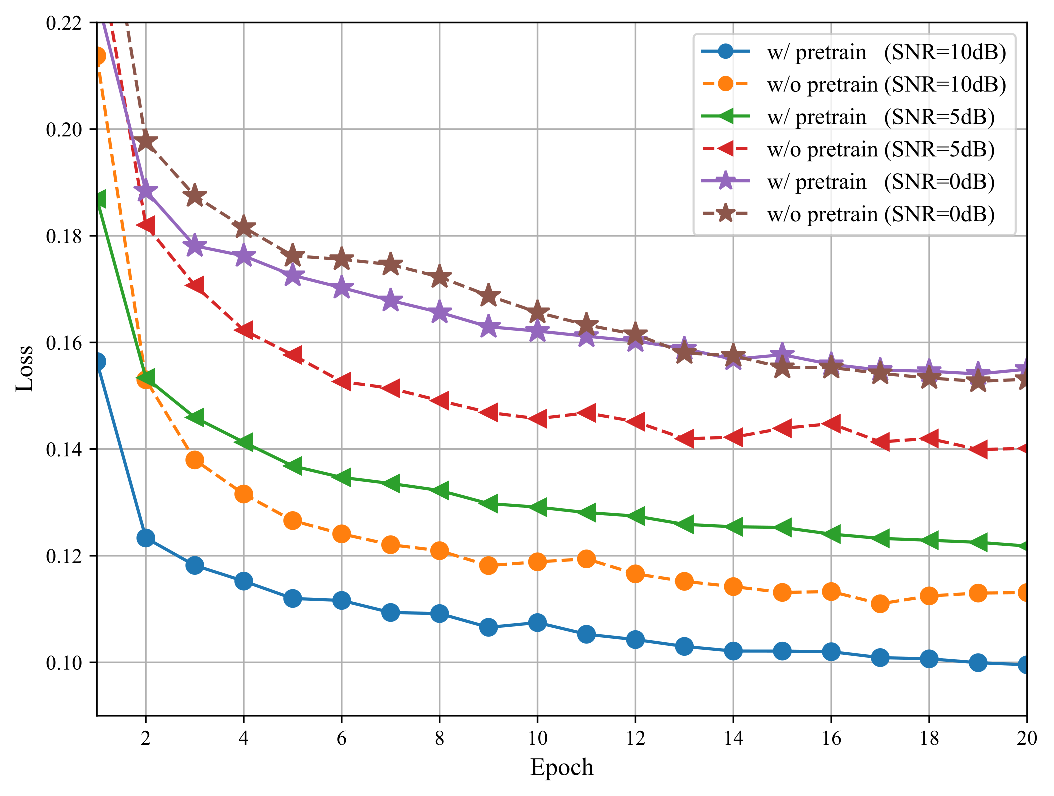} 
    \caption{Loss values versus training epochs for transfer learning. The experiments are all conducted under AWGN channel.}
    \label{img:pretrain} 
    \vspace{-0.25cm}
\end{figure}

\subsection{Transfer Learning}
In this experiment, we verify the performance of transfer learning aided LPC-FT for retraining under different channel conditions. As mentioned in \ref{sec:training}, we first get a pretrained LiDAR point cloud compression model without considering channel noise, and then train the whole transmission network for different SNRs based on the pretrained compression model. Fig. \ref{img:pretrain} shows the relationship between the training loss value and the epoch for transfer learning. The pretrained compression model provides additional knowledge about LiDAR point cloud characteristics, so its training loss is always lower than that of training from scratch for different SNRs. The performance improvement from transfer learning is more obvious for larger SNRs because the model can more effectively leverage compression knowledge without expending significant effort to counteract wireless noise. Conversely, at lower SNRs, the training loss gradually decreases to a similar value, as the model needs to adjust more parameters to effectively combat wireless noise. The results demonstrate that the shared compression model can help the network converge faster for different SNRs through transfer learning, and reduce much training costs for dynamic environments.   

\vspace{0.0cm}

\section{Conclusions}
LiDAR point clouds are crucial for multi-agents collaborative perception, but their large data volume limits system performance and efficiency. This paper presents an effective LiDAR point cloud feature transmission system (LPC-FT) for low-interaction collaborative perception among multiple agents. We first utilize a density-aware feature encoder with additional position encoding to improve compression performance. To mitigate the effects of wireless channel, we propose a channel encoder network and a feature fusion network based on the attention mechanism, ensuring robust reconstruction performance. We integrate the nonlinear activation layer for compatibility with digital communication systems and employ transfer learning to enhance network training. Experimental results show that our methods achieve outstanding reconstruction performance with reduced data costs compared with both learning-based methods and traditional methods, indicating the success of our system design.

\bibliographystyle{IEEEtran}
\bibliography{IEEEabrv,ref.bib}

\vfill

\end{document}